\documentclass[prb,twocolumn,showpacs]{revtex4}
\usepackage{graphicx}
\usepackage[T1]{fontenc}
\usepackage{amssymb,amsmath}
\begin{document}

\title{      Slave-boson approach to the metallic stripe phases 
             with large unit cells }

\author{     Marcin Raczkowski }
\affiliation{Marian Smoluchowski Institute of Physics, Jagellonian
             University, Reymonta 4, PL-30059 Krak\'ow, Poland \\
             and Laboratoire CRISMAT, UMR CNRS--ENSICAEN(ISMRA) 6508, 
             6 Bld. du Mar\'echal Juin, F-14050 Caen, France}

\author{     Raymond Fr\'esard }
\affiliation{Laboratoire CRISMAT, UMR CNRS--ENSICAEN(ISMRA) 6508, 
             6 Bld. du Mar\'echal Juin, F-14050 Caen, France}

\author{     Andrzej M. Ole\'s }
\affiliation{Marian Smoluchowski Institute of Physics, Jagellonian
             University, Reymonta 4, PL-30059 Krak\'ow, Poland  \\
             and Max-Planck-Institut f\"ur Festk\"orperforschung,
             Heisenbergstrasse 1, D-70569 Stuttgart, Germany }

\date{10 February 2006, 
      published 26 May 2006 in: Phys. Rev. B {\bf 73}, 174525 (2006)}

\begin{abstract}
Using a rotationally invariant version of the slave-boson approach in 
spin space we analyze the stability of stripe phases with large unit 
cells in the two-dimensional Hubbard model. This approach allows one
to treat strong electron correlations in the stripe phases relevant 
in the low doping regime, and gives results representative of the 
thermodynamic limit. Thereby we resolve the longstanding controversy 
concerning the role played by the kinetic energy in stripe phases.  
While the transverse hopping across the domain walls yields the largest 
kinetic energy gain in the case of the insulating stripes with one hole 
per site, the holes propagating along the domain walls stabilize the 
metallic vertical stripes with one hole per two sites, as observed in 
the cuprates. We also show that a finite next-nearest neighbor hopping 
$t'$ can tip the energy balance between the filled diagonal and 
half-filled vertical stripes, which might explain a change in the 
spatial orientation of stripes observed in the high $T_c$ cuprates 
at the doping $x\simeq 1/16$.
\end{abstract}

\pacs{74.72.-h, 71.10.Fd, 71.45.Lr, 75.10.Lp}
\maketitle

\section{Introduction}
\label{sec:1} 

The experimentally established existence of nanoscale charge order
phenomena in transition metal oxides continues to attract much 
attention. In particular, phase separation, manifesting itself in 
formation of nonmagnetic one-dimensional (1D) domain walls which 
separate the antiferromagnetic (AF) domains with opposite phases, has 
been proposed to be responsible for unusual properties of the high $T_c$ 
copper oxide superconductors.\cite{Kiv03} Moreover, taking into account 
that magnetic excitation spectra have been found to be similar among 
various high $T_c$ cuprates, the knowledge of the spatial distribution 
of holes in the CuO$_2$ planes might be a first step towards microscopic 
understanding mechanism of the superconductivity itself.\cite{Lee06} 
Indeed, the most prominent feature of the spin excitations in 
YBa$_{2}$Cu$_{3}$O$_{6+\delta}$ (YBCO) is the resonance that occurs 
at the AF wavevector ${\bf Q}=(\pi,\pi)$ and at the energy of 41 meV in 
the case of optimal doping while for lower energies incommensurate (IC) 
magnetic peaks emerge. \cite{Hay04} In contrast, for a long time it was 
thought that La$_{2-x}$Sr$_x$CuO$_4$ (LSCO) does not show the resonance 
and that the corresponding IC fluctuations are dispersionless. However, 
recent high-resolution neutron scattering studies on optimally doped 
sample have revealed that in fact the magnetic excitations are 
dispersive and quite similar to those of YBCO. \cite{Chr04} Moreover, 
the overall evolution of the magnetic scattering with energy in 
La$_{2-x}$Ba$_x$CuO$_4$ (LBCO) also resembles the one found in YBCO. 
\cite{Tra04} Altogether, the common low-energy excitations imply that 
the spin dynamics in doped copper oxide superconductors has the same 
origin. Remarkably, the observed spectra might be consistently 
explained in terms of fluctuating stripes suggesting that the stripe 
instability is a universal phenomenon of all cuprates and thus it might 
be indeed regarded as relevant for the superconductivity.\cite{Sei05}

Historically, the first compelling evidence for both magnetic and charge 
order in the cuprates, was accomplished in a neodymium codoped compound 
La$_{1.6-x}$Nd$_{0.4}$Sr$_x$CuO$_4$ (Nd-LSCO). Indeed, around hole 
doping $x=1/8$, Tranquada {\it et~al.} \cite{Tra95Nd} found that 
\emph{elastic} neutron scattering is characterized by magnetic peaks 
at the wave vectors ${\bf Q}=\pi(1\pm 2\epsilon,1)$ 
and ${\bf Q}=\pi(1,1\pm 2\epsilon)$. Such positions of the peaks 
corresponds to equally probable modulations along one of two equivalent 
lattice directions $x$ and $y$, to which we refer as horizontal and 
vertical stripes, respectively. Moreover, inspired by the pioneering 
Hartree-Fock (HF) studies suggesting that the staggered magnetization 
undergoes a phase shift of $\pi$ at the charge domain wall (DW),
\cite{Zaa89} the authors found additional Bragg peaks displaced 
symmetrically by $\pm4\pi\epsilon$ around the $\Gamma=(0,0)$ point, 
precisely at the position expected for charge modulation. Recently, 
the same geometry for static IC magnetic and charge peaks has been 
reported\cite{Fuj04} in LBCO with $x=1/8$. In fact, neutrons do not 
detect charge directly but instead they are sensitive to nuclear 
displacements induced by the charge modulation. Therefore, a very recent 
resonant soft X-ray scattering study of the charge order in LBCO at the 
same doping is of a particular great value being a more direct evidence 
of charge modulation. These studies have yielded an estimate for the 
number of holes per DW to be 0.59, very close to the expected from 
neutron scattering experiments value 0.5. It corresponds to the 
so-called \textit{half-filled} stripes, as they are characterized by the 
filling of one doped hole per two atoms along the DWs. \cite{Abb05}
Remarkably, at this particular doping, both compounds exhibit a deep 
minimum in the doping dependence of $T_c$, suggesting a strong 
competition between the \emph{static} stripe order and the 
superconductivity.\cite{Moo88}

Conversely, the large drop in $T_c$ at $x=1/8$ is not observed in LSCO, 
even though \emph{inelastic} neutron scattering experiments in the 
superconducting regime of $x\ge0.06$ have revealed the presence of 
magnetic peaks at the same IC positions as in Nd-LSCO and LBCO. 
Moreover, Yamada {\it et~al.}\cite{Yam98} established a remarkably 
simple relation $\epsilon\simeq x$ for $0.06\le x\le 0.125$ with a 
lock-in effect at $\epsilon\simeq 1/8$ for larger $x$. This means that 
increasing doping reduces the distance $d=1/2x$ between the half-filled 
stripes. Hence, it appears that such a stripe order is compatible with 
the superconductivity, which is particularly pronounced when the spin 
correlations remain purely dynamic. However, the superconducting state 
may also coexist with static stripe order,\cite{Tra97} but then the 
value of $T_c$ is strongly reduced. This conjecture is strongly 
supported by \emph{inelastic\/} neutron scattering experiments on YBCO, 
which have also established the presence of IC vertical/horizontal 
spin fluctuations throughout the entire superconducting regime of YBCO. 
\cite{Dai01} Moreover, for $\delta=0.35$, apart from spin fluctuations, 
Mook \textit{et al.} \cite{Moo02} have found an IC charge order 
consistent with the vertical/horizontal stripes.

In contrast, in the insulating spin-glass regime of LSCO $x< 0.06$, 
quasielastic neutron scattering experiments with the main weight at zero 
frequency demonstrate that IC magnetic peaks are located at the wave 
vector $\textbf{Q}=\pi(1\pm\sqrt{2}\epsilon,1\pm\sqrt{2}\epsilon)$. 
\cite{Wak99} This phenomenon has often been interpreted as the existence 
of \emph{static} diagonal stripes, even though no signatures of any 
charge modulation were observed. In spite of the change in spin 
modulation from the diagonal to vertical/horizontal one at the doping 
$x\simeq 0.06$, $\epsilon$ follows $x$ reasonably well over the entire 
low doping range $0.03\le x\le 0.125$. 
Moreover, the same type of the diagonal IC spin modulation has been 
discovered in Nd-LSCO at $x=0.05$, \cite{Wak01} indicating that while 
the vertical/horizontal spin-density modulation, accompanied 
by the charge modulation, might be regarded as a generic property 
of the superconducting regime, the diagonal orientation of the magnetic 
domains is common for the lightly Sr-doped insulating systems. Finally, 
in the narrow range of low doping $0.02\le x\le 0.024$, IC magnetic 
peaks in LSCO are observed at the wave vector 
$\textbf{Q}=\pi(1\pm\epsilon,1\pm\epsilon)$, \cite{Mat00} 
just as in the insulating La$_{2-x}$Sr$_x$NiO$_4$ (LSNO) compound. 
\cite{Sac95} In this case, the spacing between stripes is 
equal to $d=1/x$; such structures with the filling of one doped hole 
per one atom in the DW correspond to the so-called {\it filled\/} 
stripes. 

The intriguing competition between vertical and diagonal stripes 
has been already noticed in the early mean-field studies. \cite{Zaa89} 
Indeed, it has been found that the vertical stripes were favored 
for $U/t\lesssim 4$ whereas the diagonal DWs were formed at a stronger 
Coulomb repulsion. Unfortunately, these studies predicted the filled 
stripes in the ground state. 
A general feature of such instability is a gap/pseudogap precisely 
on the Fermi surface. As a result, charge transport is not possible 
in idealized filled stripes. In contrast, the ground state energy 
half-filled stripes observed in the cuprates, have been found in a few 
methods which go beyond the Hartree-Fock approximation (HFA), such as: 
density matrix renormalization group,\cite{Whi98} 
variational local ansatz approximation,\cite{Gor99} 
Exact Diagonalization (ED) of finite clusters, \cite{Toh99} 
analytical approaches based on variational trial wave 
function within the string picture, \cite{Che02,Wro00} 
dynamical mean field theory, \cite{Fle00}  
cluster perturbation theory, \cite{Zac00} and Quantum Monte Carlo (QMC) 
simulations, \cite{Bec01} indicating the crucial role of local electron 
correlations in stabilizing these structures. Indeed, while the HF 
method is well suited to compare the energies of different types of 
magnetically ordered phases in the limit of large on-site Coulomb 
interaction $U\to\infty$, where it gives the same asymptotic behavior 
as the SBA or the Gutzwiller ansatz,\cite{Ole89} it is crucial to use 
a better approach than the HFA at intermediate $U$, particularly 
when magnetic polarization of some sites is missing.

A good starting point for a proper approximate treatment of strong 
correlations is the representation of the Hubbard model in terms of 
auxiliary fermions and bosons. \cite{Kot86} 
The slave-boson approximation (SBA) has been applied successfully to a 
whole range of problems and is known to provide a realistic mean-field 
description of strongly correlated systems. 
It is quite encouraging that the ground state phase diagram obtained 
using spin-rotation-invariant (SRI) slave-boson representation of 
the two-dimensional (2D) Hubbard model with homogeneous spiral, 
AF order, ferromagnetic (FM), and paramagnetic (PM) phases shows a good 
agreement both with QMC simulations and the ED method. \cite{Fre92sp} 
The slave-boson (SB) method was also used to investigate magnetic and 
charge correlations of the $t$-$t'$-U model, \cite{Fre98} 
the ground state of the Anderson lattice model, 
\cite{Mol93} and systems with orbital degeneracy. 
\cite{Has97} Moreover, the unrestricted 
slave-boson formalism has turned out to be a powerful tool in description  
of inhomogeneous states, even though in the absence of a finite 
long-range Coulomb repulsion, completely filled diagonal stripes (FDS) 
were found to be more stable than the half-filled vertical stripes (HVS), 
suggesting that a pure Hubbard model might be insufficient to capture 
the physics in the cuprates. \cite{Sei98,Sei98V} 

Another possible extension could involve an inclusion of the 
next-nearest neighbor hopping term $t'$. There are several experimental 
and theoretical studies suggesting the presence of a finite $t'$ 
in the cuprates. Indeed, topology of the Fermi surface 
seen by angle-resolved photoemission spectroscopy \cite{Dam03} and the 
asymmetry of the phase diagrams of the hole- and electron-doped cuprates 
can be understood only by introducing $t'$. \cite{Toh04} It also offers 
an explanation for the variation of $T_c$ among different families of 
hole-doped cuprates.\cite{Pav01,Pre05} Moreover, ED studies have shown 
that while the $d$-wave superconductivity correlation is slightly 
suppressed by $t'$ in underdoped regions, it is substantially enhanced 
in the optimally doped and overdoped regions, indicating that 
$t'$ is of great importance for the pairing instability.\cite{Shi04} 

Finally, Seibold and Lorenzana have found\cite{Sei04} that a finite 
$t'$ strongly affects the optimal filling of the vertical stripes 
and consequently favors partially filled configurations in the SBA. 
Unfortunately, restrictions due to the cluster size, have not let the 
authors to reach an unambiguous conclusion concerning a crossover 
towards the FDS at low doping. Motivated by this result, we have 
investigated recently the relative stability between the latter and the
HVS with the symmetry lowered by a period quadrupling due to on-wall 
spin-density wave. Using the self-consistent HFA, one finds that a 
negative ratio of the next-to nearest neighbor hopping ($t'/t<0$) expels 
holes from the AF domains and reinforces the stripe order. Therefore, 
the half-filled stripes not only accommodate holes but also redistribute 
them so that the kinetic energy is gained, and these stripes take over 
in the regime of $t'/t\simeq -0.3$ appropriate for YBCO. \cite{Rac06} 

The purpose of this paper it to investigate the stability of stripe 
phases at low doping and to resolve the longstanding controversy 
concerning the role of the kinetic energy in the formation of stripes.
Domain walls can be viewed as topological defects in the AF phase, 
which confine the holes to 1D regions, while a crossover to a 2D 
system occurs when the stripes melt.\cite{Cas01} This leads in a 
natural way to the picture of 1D composite excitations,\cite{Che02}
which contribute to the kinetic energy by the hole motion {\it along\/}
DWs. In contrast, in the early HF studies it was demonstrated that the
stripes are stabilized by the kinetic energy increments due to the bonds
connecting the domain wall with the neighboring sites of AF domains.
\cite{Zaa96} 

The rest of the paper is as follows. In Sec.~\ref{sec:2} we shortly 
review the SRI slave-boson representation of the Hubbard model and 
introduce reciprocal space formalism, based on the stripe periodicity, 
which provides a possibility to perform calculations on larger 
($128\times$128) clusters than those studied in Ref. \onlinecite{Sei04}. 
Therefore, our approach eliminates to a large extent the role of 
finite-size effects, as we show it in Sec. \ref{sec:2d}. Further, in 
Sec.~\ref{sec:3} we investigate the stability of the already mentioned 
FDS and HVS as well as of the filled vertical stripes (FVS) and 
half-filled diagonal stripes (HDS) at the two representative doping 
levels $x=1/8$ and $x=1/16$. For stable structures we compare the 
charge, spin, and double occupancy profiles obtained in the SBA with the 
ones obtained in the HFA. We also analyze in detail kinetic and magnetic 
energy contribution as well as discuss the shape of the density of 
states (DOS). The effect of a finite next-nearest neighbor hopping $t'$ 
on the stripe stability and induced by it conspicuous changes in the 
band structure are presented in Sec. \ref{sec:4}. Finally, in Sec. 
\ref{sec:5} we present a short summary and general conclusions. 

\section{ Formalism}
\label{sec:2}

\subsection{Rotationally invariant slave-boson method} 
\label{sec:2a}

In this paper we study the 2D Hubbard model, 
\begin{equation}
H=-\sum_{ij,\sigma}t_{ij}
   c^{\dag}_{i\sigma}c^{}_{j\sigma} +
   U\sum_{i}n^{}_{i\uparrow}n^{}_{i\downarrow},
\label{eq:Hubb0}
\end{equation}
where the electron hopping $t_{ij}$ is $t$ on the bonds connecting the 
nearest neighbor sites $\langle ij\rangle$ and $t'$ between 
next-nearest neighbor sites, while $U$ stands for the on-site Coulomb 
interaction. Our study is based on the spin-rotation invariant 
formulation of the Kotliar and Ruckenstein\cite{Kot86} SB representation 
of the Hubbard model,\cite{Fre92,Fre01} in which one enlarges the 
Hilbert space of the Hamiltonian (\ref{eq:Hubb0}) by introducing 
auxiliary boson operators: $e_i$ and $d_i$, as well as a scalar 
$p^{}_{i0}$ and a vector 
$\textbf{p}^{}_i=(p^{}_{i1},p^{}_{i2},p^{}_{i3})$ bosonic field at each
site. The former operators act as projection operators on empty and 
doubly occupied sites, while the latter describe the spin and charge 
degrees of freedom in the singly occupied subspace.

The main advantage of such a representation is that the actual electron
configuration is controlled by the bosons and one can thus write the 
Hubbard interaction as a bosonic occupation number operator. In contrast, 
the operator for the kinetic energy becomes more involved since the 
motion of a physical electron changes boson occupation numbers on both 
lattice sites involved in the hopping process. Hence, in terms of the 
above SB operators, the Hubbard model (\ref{eq:Hubb0}) takes the form,
\begin{equation}
H_{SB}=-\sum_{ij}\sum_{\sigma\sigma'\sigma''}t_{ij}
   z^{\dag}_{i\sigma\sigma'}f^{\dag}_{i\sigma'} 
   f^{}_{j\sigma''}z^{}_{j\sigma''\sigma} 
  + U\sum_{i}d^{\dag}_{i}d^{}_{i},
\label{eq:Hubb_sb}
\end{equation}
where $\underline{z}_i$ are $2\times 2$ matrices in spin space which 
depend on the actual configuration of the boson fields, as explained 
in Ref. \onlinecite{Zim97}. These matrices are constructed in such a 
way that the correct mean-field result in the noninteracting limit 
(at $U=0$) is recovered. Additionally, the SB operators have to 
fulfill the following constraints at each site, 
\begin{equation}
e_i^{\dag}e_i^{}  + d_i^{\dag}d_i^{} 
                  + \sum_{\mu} p_{i\mu}^{\dag}p_{i\mu}^{} = 1, 
\label{eq:const1} 
\end{equation}
\begin{equation}
2d_i^{\dag}d_i^{} + \sum_{\mu} p_{i\mu}^{\dag}p_{i\mu}^{} = 
\sum_{\sigma}f_{i\sigma}^{\dag}f_{i\sigma}^{}, 
\label{eq:const2} 
\end{equation}
\begin{equation}
p_{0i}^{\dag}\textbf{p}_i^{} + \textbf{p}_i^{\dag}p_{0i}^{}
                  - i\textbf{p}_i^{\dag}\times\textbf{p}_i^{} =
\sum_{\sigma\sigma'}\boldsymbol{\tau}_{\sigma\sigma'}^{}
f_{i\sigma'}^{\dag}f_{i\sigma}^{}. 
\label{eq:const3}
\end{equation}
They are enforced by corresponding Lagrange multipliers in the 
action, and one obtains a path integral representation of the Hubbard
model. Owing to the gauge symmetry group of the action, the phase of
five of the six bosonic fields can be gauged away.\cite{Fre92} These
fields can be handled in the Cartesian gauge, or in the radial gauge.
\cite{Fre01}In the following we handle the action in the Cartesian 
gauge at saddle-point level (in the SBA). This procedure is exact in 
the large degeneracy limit.\cite{Fre92,Flo02} Comparison of ground 
state energies with data resulting from numerical simulations show very 
good agreement.\cite{Fre91,Fre92sp} 

Since homogeneous AF phases turned out to be unstable with respect to
spiral phases,\cite{Fre91} one may anticipate that a similar effect
will take place in stripe phases. Therefore, the present rotationally 
invariant formalism opens a possibility of studying also more general 
structures which we postpone to future studies. Here we will limit 
ourselves to noncanted states which are locally stable, as checked 
in a few representative cases. 

\subsection{ Reciprocal space approach for stripe phases }
\label{sec:2b}

\begin{figure}[t!]
\begin{center}
\unitlength=0.01\textwidth
\begin{picture}(50,55)
\put(11,31){\includegraphics[width=4.1cm ]{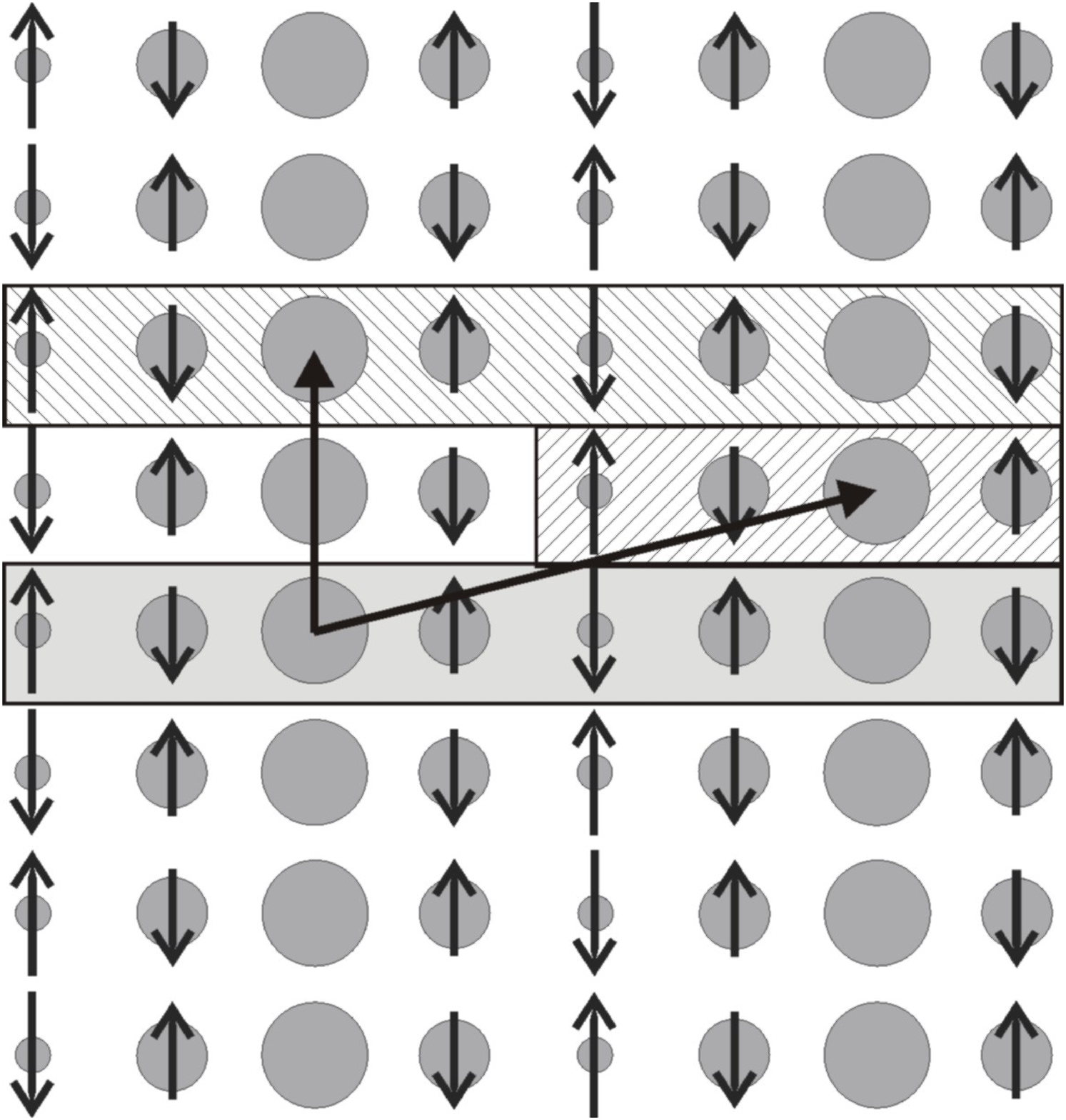}}
\put(11, 5){\includegraphics[width=4.1cm ]{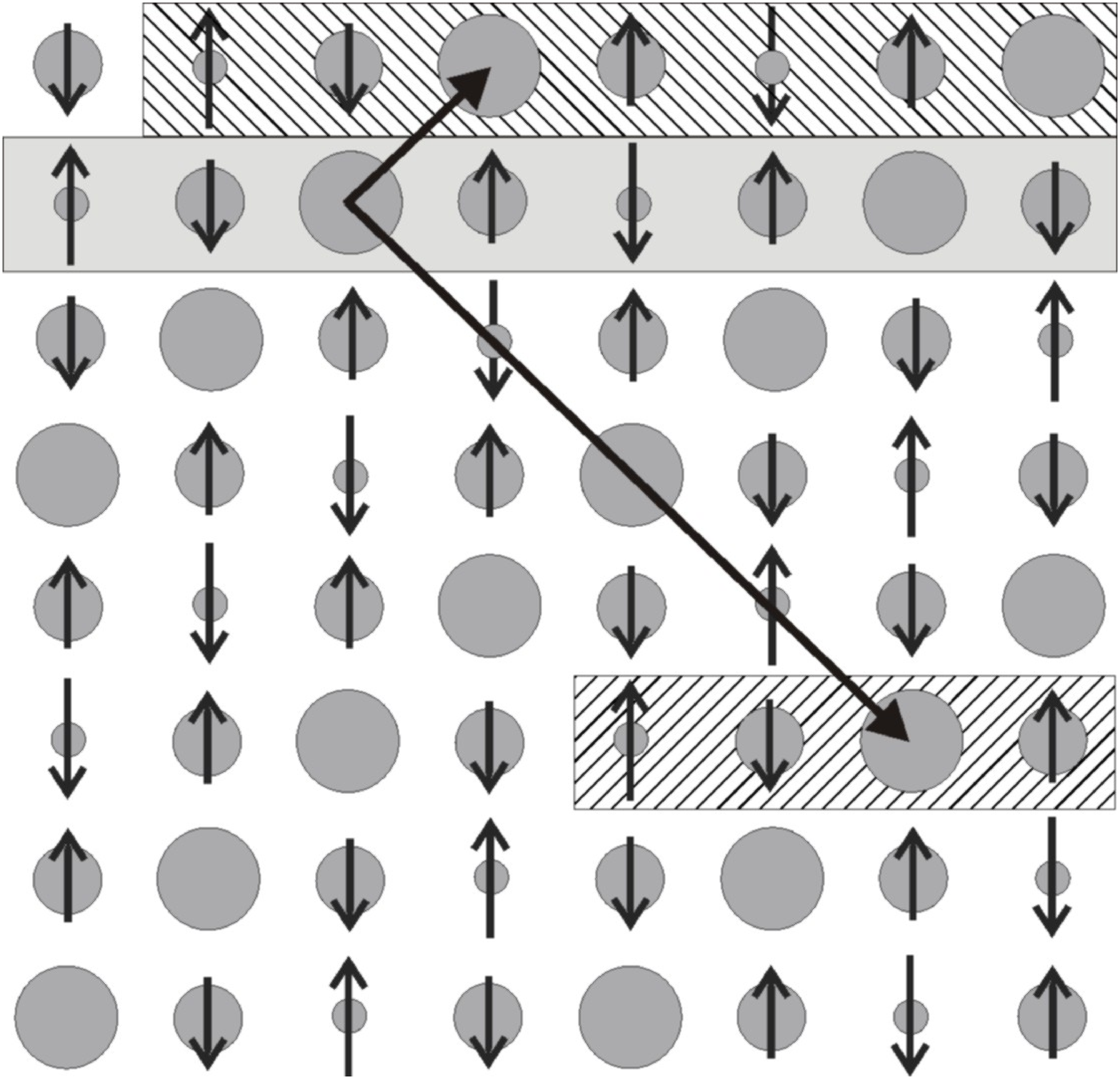}}
\put(7,1){\vector(1,0){10}}
\put(7,1){\vector(0,1){10}}
\put(11,2) {\large{$l_x$} }
\put(4.5,5){\large{$l_y$} }
\put(3,53){ {\Large (a)} }
\put(3,25){ {\Large (b)} }
\end{picture}
\end{center}
\caption{
Symmetry properties of stripe phases:
(a) the vertical stripe phase --- its unit cell with two periodicity 
vectors: $\textbf{g}_1=\bigl(4,1\bigr)$ and 
$\textbf{g}_2=\bigl(0,2\bigr)$;
(b) the diagonal stripe phase --- its unit cell with two periodicity 
vectors: $\textbf{g}_1=\bigl(1,1\bigr)$ and 
$\textbf{g}_2=\bigl(4,-4\bigr)$.
}
\label{fig:unit}
\end{figure}

In order to obtain unbiased results for stripe phases one should carry 
out calculations on large clusters. So far, such calculations were 
usually performed in real space to address directly inequivalent 
sites in the stripe unit cell. However, an efficient approach may be 
constructed by making use of the periodicity of a stripe phase which 
allows one to cover the entire lattice by small unit cells. Indeed, 
each position vector $\textbf{R}_i$ of an arbitrary atom can be
decomposed into a sum involving certain periodicity vectors  
$\textbf{g}_1$ and $\textbf{g}_2$, and a vector
$\boldsymbol{\delta}_{M}$ labeling all  
inequivalent sites $i$ within the unit cell: 
\begin{equation}
\textbf{R}_i = n\textbf{g}_1+m\textbf{g}_2+\boldsymbol{\delta}_{M},
\label{eq:Ri}
\end{equation} 
with integer $\{n,m\}$. 
Accordingly the covering of the Brillouin zone spanned by the
wave vector $\textbf{k}$ is achieved by the decomposition:
\begin{equation}
\textbf{k}=\textbf{K}+\textbf{G},
\label{eq:k}
\end{equation}
with $\textbf{G}$ belonging to a set of wave vectors representative of
the periodicity of the structure of interest.\cite{Asc76} For a stripe
phase with vertical domain walls separated by $d=4$ lattice spacings, as 
shown in Fig.~\ref{fig:unit}(a), the magnetic unit cell consists of a 
row of eight atoms perpendicular to the stripes. One immediately finds 
two periodicity vectors 
$\textbf{g}_1 = \bigl(4,1\bigr)$ and 
$\textbf{g}_2 = \bigl(0,2\bigr)$. In this case the set of vectors
$\textbf{G}$ satisfying $\exp\{i\textbf{g}_{(1,2)}\cdot\textbf{G}\}=1$
is given by 
\begin{equation}
\textbf{G}_1^{(l)} = 
\frac{2\pi}{8} (2l,0),
\qquad
\textbf{G}_2^{(l)} = 
\frac{2\pi}{8}\bigl(2l+1,4\bigr),
\label{eq:GVSC}
\end{equation}
with $0\le l \le 3$. 

A second example is a diagonal stripe phase depicted in Fig. 
\ref{fig:unit}(b). Again, considering the periodicity of this phase,
one finds the magnetic unit cell consisting of a row of eight atoms 
along the $x$-direction. However, the periodicity vectors are now given 
by $\textbf{g}_1=(1,1)$ and $\textbf{g}_2=(4,-4)$. Accordingly one
finds for $\textbf{G}$:
\begin{equation}
\textbf{G}_1 = (0,0),
\qquad
\textbf{G}_2^{(l)} = 
\frac{2\pi}{8}(l,8-l),
\label{eq:DVSC}
\end{equation}
with $1\le l \le 7$. 

In both cases the vectors $\textbf{K}$ in Eq. (\ref{eq:k}) are spanning 
the rectangle $[0,2\pi/8]\times [0,2\pi]$. The extension to other 
stripe phases is straightforward. Using the above procedure, the 
fermionic contribution to the action in the momentum representation is
given by: 
\begin{equation}
{\mathcal S}_f \!=\! \int_0^{\beta}\! d \tau
\sum_{\textbf{K}\textbf{G}\textbf{G}'}
\sum_{\sigma\sigma'}
f_{\textbf{K}+\textbf{G},\sigma}^{\dag} (\tau)
M_{\textbf{GG}'}^{\sigma\sigma'}(\textbf{K},\tau)
f_{\textbf{K}+\textbf{G}',\sigma'}^{}(\tau) ,
\label{eq:fiMfj}
\end{equation} 
with,
\begin{align}
M_{\textbf{GG}'}^{\sigma\sigma'}(\textbf{K},\tau) &= (\partial_{\tau}
-\mu) \delta_{\sigma\sigma'} \delta_{\boldsymbol{G}\boldsymbol{G}'}
+\frac{1}{L} \sum_{\boldsymbol{\delta}_M}
e^{-i(\boldsymbol{G}-\boldsymbol{G}')\boldsymbol{\delta}_M} \nonumber \\
&\hskip -1cm \times\Bigl[\beta_{0\boldsymbol{\delta}_{M}}(\tau)
\delta_{\sigma\sigma'}
 + \boldsymbol{\beta}_{\boldsymbol{\delta}_{M}}(\tau)
   \cdot\boldsymbol{\tau}_{\sigma\sigma' }                  \nonumber \\
&\hskip -1cm -\sum_{\boldsymbol{\delta}}
t_{\boldsymbol{\delta}_M,\boldsymbol{\delta}_M+\boldsymbol{\delta}} 
e^{i(\boldsymbol{K}+\boldsymbol{G}')\boldsymbol{\delta}} 
\left (z_{\boldsymbol{\delta}_M+\boldsymbol{\delta}}^{}(\tau)
z_{\boldsymbol{\delta}_M}^{\dag}(\tau)\right)^T_{\sigma\sigma'}\Bigl].  
\label{eq:M_sb}
\end{align}
Here $L$ is the number of atoms in the unit cell so that the ratio 
$N/L$ gives the number of unit cells needed to cover a whole cluster 
with $N$ sites, $\partial_{\tau}$ is an imaginary time derivative, 
and $\mu$ is the chemical potential. The quantities $\beta_{0}$ and
$\boldsymbol{\beta}$ are the Lagrange multipliers enforcing the 
constraints given in Eq. (\ref{eq:const2}) and Eq. (\ref{eq:const3}),
respectively.

Hence, we have reduced the fermionic matrix 
$M_{\textbf{kk}'}^{\sigma\sigma'}$ down to decoupled blocks labeled by 
$\textbf{K}$. This procedure gives a considerable time gain during 
numerical calculations. While the usual workload in a diagonalization 
algorithm is $\sim N^3$ (see Ref. \onlinecite{Pre96}), the number of 
operations needed for diagonalization of $N/L$ smaller matrices is only 
$\sim\frac{N}{L}\cdot L^3=NL^2$. This means that the symmetry reduction 
makes a tremendous simplification and the time needed for numerical 
calculations is reduced by a factor $\left(N/L\right)^2$ as compared to 
a straightforward 'brute force' diagonalization of an $N\times N$ 
matrix, which is necessary when the computations are performed in real 
space.

\subsection{ Mean-field approximation}
\label{sec:2c}

In Secs. III and IV we present solutions for the stripe phases obtained 
within the SB mean-field approximation in which one replaces the Bose 
fields and the Lagrange multipliers by their time-independent averages. 
Hence, the free energy found at the temperature $T$ follows as,
\cite{Fre92sp}
\begin{align}
F & = \frac{1}{L} \Bigl\{ \sum_l\bigl[
      \alpha_l^{}(e^2_l+p_{0l}^2+p_l^2 + d_l^2-1) \nonumber \\
  & - \beta_{0l}^{}(p_{0l}^2+p_l^2 + 2d_l^2) +  U_{}^{}d_l^2 
    -  2\boldsymbol{\beta}_l^{}\cdot\textbf{p}_l^{}p_{0l}^{}\bigr] 
\nonumber \\ 
  & -  \frac{1}{\beta}\sum_{\textbf{q}\sigma} 
       \ln\bigl( 1+ e^{-\beta\varepsilon_{\textbf{q}\sigma}}\bigr) 
    + \mu N_{el} \Bigr\}, 
\label{eq:F_sb}
\end{align}
with $\beta=1/k_BT$, $\varepsilon_{\textbf{q}\sigma}$ standing for the
eigenmodes of the fermionic matrix (\ref{eq:M_sb}), and $N_{el}$ 
corresponding to the total number of particles. The equilibrium values 
of the classical field amplitudes are determined from the saddle-point 
equations in which the partial derivatives are taken with respect to 
three Lagrange multipliers enforcing the constraints 
(\ref{eq:const1})--(\ref{eq:const3}), as well as to the four SB 
fields $p_{0l}$, $p_l$, $e_l$, and $d_l$, for each inequivalent 
site $l=(l_x,l_y)$ within the elementary stripe unit cell. The
saddle-point equations are solved using the Powell hybrid method.

\subsection{ Finite-size effects }
\label{sec:2d}

\begin{figure}[t!]
\begin{center}
\includegraphics*[width=8.2cm ]{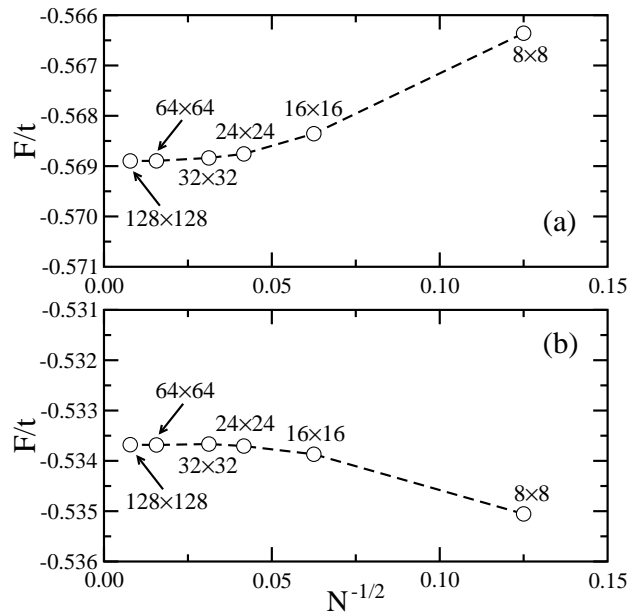}
\end{center}
\caption{
Finite-size scaling of free energy $F$ for: 
(a) HVS, and 
(b) HDS, 
as obtained within the SBA for the Hubbard model at doping $x=1/8$.
Parameters: $U=12t$ and $\beta t=1000$.
}
\label{fig:F_N}
\end{figure}

In order to estimate to what extent the calculation method eliminates 
finite-size effects, we examine the size dependence of the free energy 
at low temperature $\beta t=1000$ by considering stripe phases on 
clusters of increasing size: $8\times 8$, $16\times 16$, $24\times 24$, 
$32\times 32$, $64\times 64$, and $128\times 128$. All systems are 
described by the Hubbard model with $U=12t$ and $x=1/8$. The 
finite-size scaling of the free energy obtained for the HVS obtained 
within the SBA is shown in Fig.~\ref{fig:F_N}(a). As it could be 
expected, finite size effects are particularly severe for small 
$8\times 8$ and $16\times 16$ clusters but a further increase of the 
system size results in a gradual saturation of the free energy. The 
systematic errors due to these effects are diminished to the order of 
$10^{-5}t$ when the calculations are performed on clusters with $10^{4}$ 
atoms. 

The above behavior is universal, and a similar tendency appears also 
in a system with the HDS [see Fig.~\ref{fig:F_N}(b)]. Note, however, 
that in the case of the vertical stripes increasing cluster size lowers 
the total energy. In contrast, the energy increases when the number of 
atoms in a cluster with the diagonal stripes gets larger, suggesting 
that small cluster calculations underestimate (overestimate) the free 
energy of the vertical (diagonal) stripe phase, respectively, and may 
thus lead to systematic errors and inconclusive results in some cases. 
Even though the free energy $F$ saturates quickly with an increasing 
number of sites in both cases, one should keep the above feature in 
mind, especially when it comes to investigate the relative stability of 
vertical and diagonal phases as the inclusion of quantum fluctuations 
could substantially strengthen this effect. Here we would like to point 
out another virtue of the present studies performed at temperature 
$\beta t=1000$. When the temperature is so low, the entropy contribution
to the free energy is almost entirely suppressed as 
$F\simeq\langle H\rangle$, so that one may analyze only the internal 
energy for different phases, as done below.

\section{ Electron correlation effects in stripe phases }
\label{sec:3}

\subsection{ Physical characterization of stripe phases }
\label{sec:3a}

Let us now examine the role of strong electron correlations in 
stabilizing stripe phases of various type. Even though the HFA and SBA
yield quantitatively very similar results at half-filling, as recently
reviewed by Korbel {\it et~al.},\cite{Kor03} we show below that they
differ markedly in the stripe phases. Indeed, since the HF wave 
function is a Slater determinant of single particle states, it does not 
provide enough variational freedom to implement the many-body processes 
that are relevant for the behavior of interacting fermions. Therefore, 
one should expect that the inclusion of the correlation effects modifies 
considerably the distribution of charge and spin density in a stripe 
phase, especially around nonmagnetic DWs where the correlation 
corrections are large.\cite{Gor99}   

To illustrate this point, we investigate the charge and magnetization 
distribution in the SBA and compare them with the reference states 
obtained in the HFA. To describe the charge distribution, we introduce 
the local hole density for each inequivalent site $l=(l_x,l_y)$ in the 
stripe unit cell, 
\begin{equation}
n_{\rm h}(l_x) = 1 - \sum_{\sigma}
\bigl\langle f^{\dag}_{l\sigma}f^{}_{l\sigma}\bigr\rangle,
\label{eq:nh}
\end{equation}
where the average electron density,
\begin{equation}
\langle f^{\dag}_{l\sigma}f^{}_{l\sigma}\rangle =
\sum_{\textbf{q}} \Psi_{l\sigma}^{\dag}(\textbf{q})
\Psi_{l\sigma}^{}(\textbf{q})n(\epsilon_{\textbf{q}\sigma}), 
\label{eq:n} 
\end{equation}
is given by the real space eigenvectors $\Psi_{l\sigma}^{}(\textbf{q})$
of the effective Hamiltonian (\ref{eq:fiMfj}), and
$n(\varepsilon_{\textbf{q}\sigma})$ is the Fermi-Dirac distribution.
Further, magnetic domain structures are best described by the 
modulated magnetization density, 
\begin{equation}
S_{\pi}(l_x) = (-1)^{l_x+l_y}\frac{1}{2}
\sum_{\sigma\sigma'}\bigl\langle f_{l\sigma}^{\dag} 
\boldsymbol{\tau}_{\sigma\sigma'}^{} f_{l\sigma'}^{}\bigr\rangle,
\label{eq:Spi}
\end{equation}
with a site dependent phase factor $(-1)^{l_x+l_y}$ compensating the 
modulation of the staggered magnetization density within a single AF 
domain, as well as by double occupancy, 
\begin{equation}
D(l_x) = \bigl\langle n_{l\uparrow}n_{l\downarrow}\bigr\rangle.
\label{eq:D}
\end{equation} 

\subsection{ Stripes at low doping $x=1/16$ }
\label{sec:3b}

\begin{figure}[t!]
\begin{center}
\includegraphics*[width=8.2cm ]{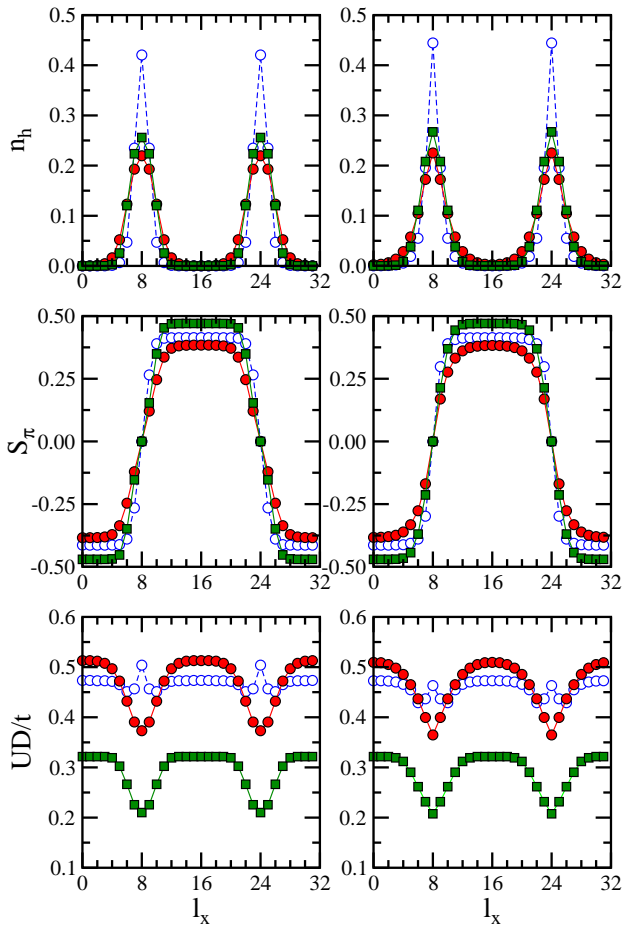}
\end{center}
\caption
{(Color online)
Hole and magnetization distribution in filled stripe phases at low 
doping $x=1/16$:
local hole $n^{}_{\rm h}(l_x)$ density (top),
magnetization $S_{\pi}(l_x)$ density (second row), and 
double occupancy $D(l_x)$ (bottom), 
for the phase with the FVS (left) and FDS (right), found 
in the Hubbard model on a $128\times 128$ cluster at $\beta t=1000$. 
Open (filled) circles show the HFA (SBA) results for $U=6t$, respectively, 
while filled squares denotes the SBA data for $U=12t$. 
}
\label{fig:nszdF16}
\end{figure}

We begin with the detailed analysis of the charge and magnetization 
distribution in the stripe phases at low doping $x=1/16$, and compare
the results of the SB treatment with those obtained in the HFA. 
The corresponding SB profiles of the FVS (left) and FDS (right) found 
in the Hubbard model with $U=6t$ at the doping $x=1/16$ are shown in 
Fig.~\ref{fig:nszdF16} (filled circles). In this case nonmagnetic DWs 
are separated by $d=16$ lattice constants so that the charge 
(magnetic) unit cell contains sixteen (thirty two) atoms, respectively. 
In agreement with the calculations of Ref.~\onlinecite{Sei98}, we note 
that the hole density $n_h$ at nonmagnetic DWs is reduced nearly twice 
in the SBA as compared to the corresponding HFA value (open circles), 
regardless of the stripe direction. Moreover, the HF stripe phases are 
characterized by the enhanced spin polarization of the AF domains. 
Such a strong modification follows directly from the fact that in the 
HFA there are only two straightforward possibilities which allow one 
to diminish the energy of the stripe phase due to the on-site Coulomb 
repulsion,
\begin{equation}
E_U = U\frac{1}{L}\sum_{l_x}D(l_x).
\label{eq:E_U}
\end{equation}
The simplest way of keeping apart electrons with the opposite spins is 
by creating strong spin polarization. As a consequence, a well known 
feature of the HFA is that it overestimates by far the tendency towards 
symmetry broken states. Another way of reducing $E_U$ which may be 
realized for inhomogeneous charge distribution, for instance in stripe 
phases, is to suppress locally the total electron density at the sites 
with small or vanishing magnetization. 

In contrast, the SB approach implements local correlations by offering 
an important mechanism to optimize the on-site interaction by an 
additional variational parameter, i.e., the boson field $d_i$ at each
site. In fact, the largest correction of the HF value is obtained at the
DW unpolarized sites, where the double occupancy $D$, Eq. (\ref{eq:D}), 
shows distinct minima in both phases (see Fig. \ref{fig:nszdF16}) 
which allows one to optimize $E_U$ even without a great reduction of 
the actual electron density. Simultaneously, a large value of $D$ within 
the AF domains suppresses partially the spin polarization of the atoms 
separating DWs, which enables more intersite excitations and leads to a 
more favorable kinetic energy gain. Taken together, these two effects 
are jointly responsible for smoother charge and spin density 
profiles with respect to the ones found in the HFA. Notice that both 
approximations yield narrow \emph{diagonal\/} stripes, revealing their 
more localized character as compared to the vertical ones and hence 
a more favorable average on-site energy, which stabilizes them in 
the strong coupling regime, as discussed below. 

\begin{figure}[t!]
\begin{center}
\includegraphics*[width=8.2cm ]{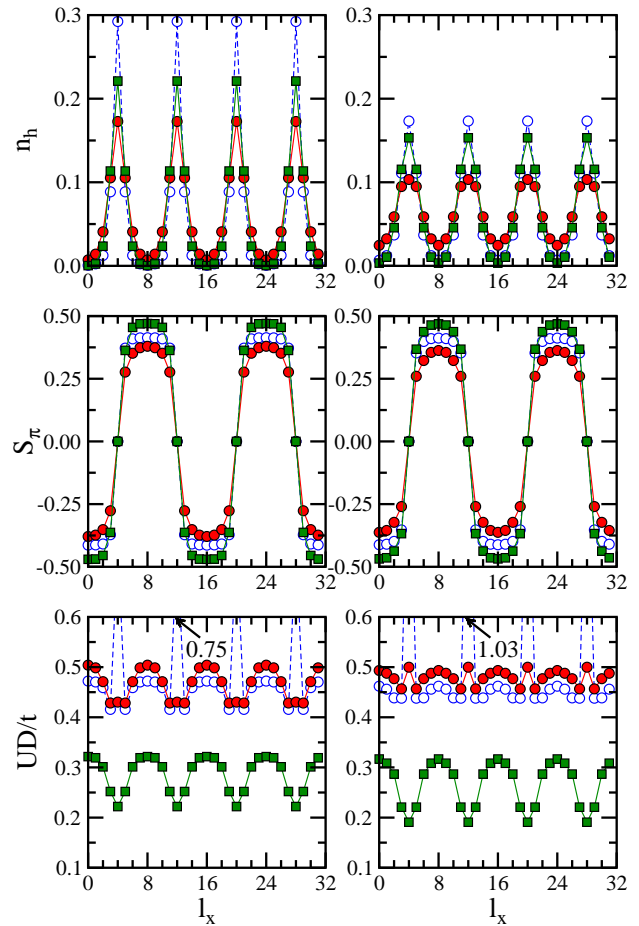}
\end{center}
\caption
{(Color online)
The same as in Fig.~\ref{fig:nszdF16} but for the half-filled stripe 
phases: HVS (left) and HDS (right). 
Open (filled) circles show the HF (SB) results for $U=6t$, respectively, 
while filled squares denotes the SB data for $U=12t$. High values of 
$UD/t>0.6$ obtained for the DW atoms in the HFA are indicated by arrows.  
}
\label{fig:nszdH16}
\end{figure}

It is quite remarkable that the present calculations performed on large 
clusters yield also locally stable half-filled stripes both in the HFA 
and in the SBA without any necessity of quadrupling of the period along 
the stripes by an additional on-wall spin-density wave.\cite{Zaa96} 
At $x=1/16$, the condition of having one doped hole per two DW atoms 
requires the charge (magnetic) unit cell containing eight (sixteen) 
atoms, respectively. The obtained profiles of the HVS (left) and HDS 
(right) are shown in Fig. \ref{fig:nszdH16}. Again, one finds a smaller 
SB charge modulation and a stronger spin polarization of the AF domains 
in the HFA. Note, however, that contrary to the case with the filled 
DWs, a narrower charge and spin profile of the \emph{vertical} stripe 
with respect to the diagonal one is apparent. 

At strong coupling with $U=12t$ (filled squares), one finds for the most 
stable phases that the double occupancy profile is interpolating between 
$D\simeq 4t^2/U^2$ deep in the magnetic domains (the value expected in 
an AF phase at half-filling\cite{Mol93b}), and the value expected in the 
PM phase at the actual hole density realized at the domain wall sites, 
as shown in Table \ref{tab:do1_16}. Thus the present approach is 
flexible enough to implement the compromise between optimized 
superexchange in the magnetic domains, and optimized (with respect to 
the local density) double occupancy in the domain walls, with, however, 
the notable exception of the HDS. In that case this optimization cannot 
be achieved, and instead one gets a more spread hole distribution. The 
same tendency is also realized at intermediate coupling.

\begin{table}[b!]
\caption 
{
Hole density $n_{\rm h}({\rm DW})$ and double occupancy $D({\rm DW})$
at domain-wall sites, and the corresponding double occupancy in the PM 
phase $D_{\rm PM}$, as found for various stripe phases in the Hubbard 
model on an $128\times 128$ cluster at $x=1/16$ doping within the SBA. 
Parameters: $U=12t$, $\beta t=1000$. 
}
\label{tab:do1_16}
\begin{ruledtabular}
\begin{tabular}{cccccc}
& \multicolumn{1}{c}{phase}                 &
\multicolumn{1}{c}{$n_{\rm h}({\rm DW})$}       &
\multicolumn{1}{c}{$D({\rm DW})$}               &
\multicolumn{1}{c}{$D_{\rm PM}$}         & \\
\colrule
 & HDS & 0.153 & 0.016 & 0.019 & \\
 & HVS & 0.221 & 0.018 & 0.018 & \\
 & FVS & 0.256 & 0.017 & 0.017 & \\
 & FDS & 0.267 & 0.017 & 0.016 & \\ 
\end{tabular}
\end{ruledtabular}
\end{table}

Experimentally, two stripe phases have been found:
(i) filled diagonal stripe phase, and 
(ii) half-filled vertical one. 
Both of them have been observed in LSCO; the former was 
found around $x=0.02$,\cite{Mat00} whereas the latter --- at a higher 
$0.06\le x\le 0.125$ doping level.\cite{Yam98} In the present 
calculations these phases were reproduced as two possible ground states 
of the doped Hubbard model. The crossover found experimentally 
indicates that the filling and orientation of DWs are indeed closely 
related in the cuprates. In other words, \emph{half-filled\/} stripes 
tend to be aligned vertically/horizontally, whereas \emph{filled\/} 
ones gain more energy by the diagonal arrangement.

The above important property can also be deduced from the free energies 
$F$ calculated in the SBA from Eq.~(\ref{eq:F_sb}) for all four stripe 
phases, given in Table~\ref{tab:Ftp0x1_16}. For completeness, the 
energies of stripe phases are compared with those of the uniform AF 
phase. Note that the sequence of stripe phases, ordered according to 
their decreasing energy, is precisely the same in both methods and one 
does find that the half-filled vertical stripe phase is preferred over 
its diagonal counterpart, and also that the filled diagonal stripe 
structure is promoted over the corresponding vertical stripe phase. 
However, noticeable discrepancies emerge when it comes to compare the 
on-site Coulomb energy $E_U$ and the nearest neighbor kinetic energy
determined in SBA, 
\begin{equation}
E_t = -t\,\frac{1}{L}\sum_{\langle l,l'\rangle}\sum_{\sigma\sigma'} 
\bigl(z^{}_{l'}z^{\dag}_{l}\bigr)^{T}_{\sigma'\sigma}\bigl\langle 
f^{\dag}_{l\sigma} f^{}_{l'\sigma'} \bigr\rangle, 
\label{eq:ekt}
\end{equation}
with the ones obtained in the HFA. Most importantly, the inclusion of 
electron correlations reduces $E_U$ of the PM phase by a factor two 
without a strong suppression of $E_t$, which improves significantly its 
free energy. Therefore, as the average hole density at the half-filled 
DWs is noticeably smaller than at the filled ones (\textit{cf}. Figs. 
\ref{fig:nszdF16} and \ref{fig:nszdH16}), it should also help to 
stabilize half-filled stripes, because one gains more correlation energy 
when the nonmagnetic atoms are close to half-filling. These features are 
clearly seen in Table~\ref{tab:Ftp0x1_16}. By comparing the repulsion 
energy, one finds that $E_U$ of both half-filled stripe phases is 
reduced in the SBA, while $E_U$ of the filled ones is even slightly 
enhanced. As a result, for moderate coupling, the structure with the HVS 
in the SBA is characterized by the smallest net double occupancy and 
hence by the most favorable $E_U$. Nevertheless, in the SB ground state 
one recovers the phase with the FDS, as a small energy loss due to the 
increased on-site energy is easily overcompensated by the kinetic energy 
gain. The FDS are recovered at strong coupling as well, where the HVS
are not even favored by $E_U$.

\begin{table}[b!]
\caption 
{
Comparison of the site-normalized free energy $F$, on-site Coulomb 
energy $E_U$, and kinetic energy $E_t$, as obtained for various phases 
in the Hubbard model within the HFA (for $U=6t$) and SBA (for $U=6t$ and 
$12t$) on an $128\times 128$ cluster at doping $x=1/16$ and 
$\beta t=1000$. For clarity, the phases are listed according to their 
decreasing energy from top to bottom. 
}
\label{tab:Ftp0x1_16}
\begin{ruledtabular}
\begin{tabular}{cccccc}
\multicolumn{1}{c}{method}        &
\multicolumn{1}{c}{$U/t$}         &
\multicolumn{1}{c}{phase}         &
\multicolumn{1}{c}{$E_U/t$}       &
\multicolumn{1}{c}{$E_t/t$}       &
\multicolumn{1}{c}{$F/t$}         \\
\colrule
  HFA & 6& HDS & 0.5185 & $-$1.1080  & $-$0.5895  \\
      &  &  AF & 0.4928 & $-$1.1198  & $-$0.6270  \\
      &  & HVS & 0.4892 & $-$1.1242  & $-$0.6350  \\
      &  & FVS & 0.4690 & $-$1.1218  & $-$0.6528  \\
      &  & FDS & 0.4586 & $-$1.1154  & $-$0.6568  \\ 
  SBA & 6 & HDS & 0.4796 & $-$1.1708 & $-$0.6912  \\ 
      &  &  AF & 0.4943 & $-$1.2009 & $-$0.7066  \\
      &  & HVS & 0.4663 & $-$1.1809 & $-$0.7146  \\
      &  & FVS & 0.4709 & $-$1.1929 & $-$0.7220  \\
      &  & FDS & 0.4671 & $-$1.1902 & $-$0.7231  \\ 
  SBA &12&  AF & 0.3181 & $-$0.7355 & $-$0.4174  \\
      &  & HDS & 0.2673 & $-$0.7023 & $-$0.4350  \\ 
      &  & HVS & 0.2860 & $-$0.7356 & $-$0.4496  \\
      &  & FVS & 0.2930 & $-$0.7455 & $-$0.4525  \\
      &  & FDS & 0.2903 & $-$0.7464 & $-$0.4561  \\ 
\end{tabular}
\end{ruledtabular}
\end{table}

Finally, it should be emphasized that, in contrast to the HFA, the most 
favorable SB phase in the moderate coupling regime $U=6t$ for the 
intersite kinetic energy $E_t$ is the AF structure (see Table 
\ref{tab:Ftp0x1_16}). However, owing to a strong suppression of the 
double occupancy at nonmagnetic DWs, the existence of stripes opens a 
possibility for a charge redistribution which optimizes $E_U$ even 
better. Consequently, in the  strong coupling regime with $U=12t$ all 
the stripe phases are favored over the uniform AF order. 

\begin{figure}[t!]
\begin{center}
\includegraphics*[width=8.2cm ]{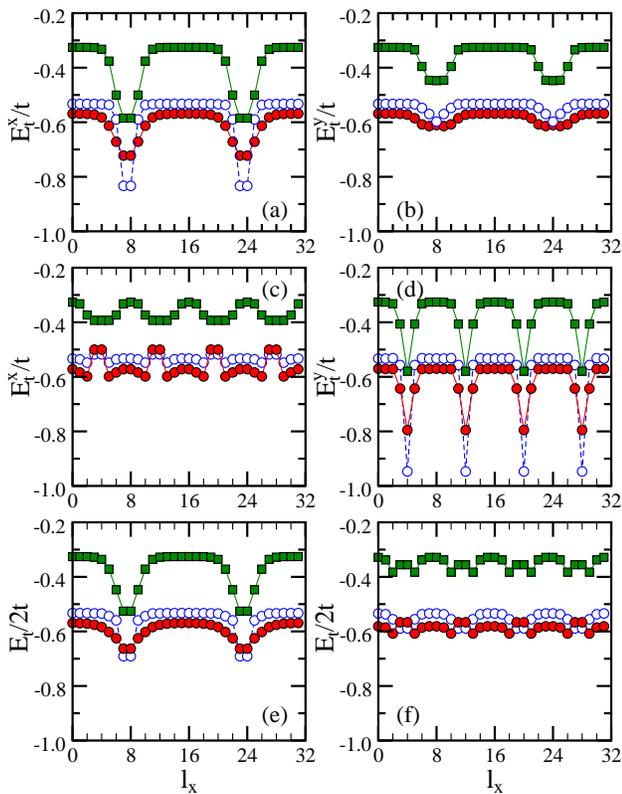}
\end{center}
\caption
{(Color online)
Kinetic energy projected on the bonds in the $x$ ($y$) direction, 
$E_{t}^{x}(l_x)$ and $E_{t}^{y}(l_x)$, as found in the phase with either 
the FVS (a,b) or HVS (c,d), shown in Figs.~\ref{fig:nszdF16} and 
\ref{fig:nszdH16}. 
Open (filled) circles show the HF (SB) results for $U=6t$, respectively, 
while filled squares denote the SB data for $U=12t$. Panels (e) and (f) 
depict the kinetic energy profiles $E_{t}/2t$ of the corresponding FDS 
and HDS.  
}
\label{fig:Ekin16}
\end{figure}

In order to identify in detail processes leading to the stripe 
stabilization, we determined the expectation values of the bond hopping 
terms, $E_t^{x}(l_x)$ and $E_t^{y}(l_x)$, along the $x$ and $y$ 
direction, respectively,
\begin{align}
E_t^x(l_x)&= -t\sum_{\sigma\sigma'} q_{\sigma\sigma'}^x(l_x)\bigl\langle 
f^{\dag}_{(l_x,l_y),\sigma} f^{}_{(l_{x+1},l_y),\sigma'} \bigr\rangle, \\
E_t^y(l_x)&= -t\sum_{\sigma\sigma'} q_{\sigma\sigma'}^y(l_x)\bigl\langle 
f^{\dag}_{(l_x,l_y),\sigma} f^{}_{(l_x,l_{y+1}),\sigma'} \bigr\rangle,
\label{eq:exy}
\end{align}
where,
\begin{align}
q_{\sigma\sigma'}^x(l_x)&=
\Bigl (z^{}_{(l_{x+1},l_y)}z^{\dag}_{(l_x,l_y)}\Bigr)^T_{\sigma'\sigma},\\
q_{\sigma\sigma'}^y(l_x)&=
\Bigl (z^{}_{(l_x,l_{y+1})}z^{\dag}_{(l_x,l_y)}\Bigr)^T_{\sigma'\sigma},
\label{eq:qxy}
\end{align} 
are the corresponding band narrowing factors. The variation of the 
kinetic energy gain across the unit cell of the phase with the FVS is 
shown in Figs. \ref{fig:Ekin16}(a) and \ref{fig:Ekin16}(b). 
Deep in the magnetic domains, for large $U$, the kinetic energy in
both directions is given by $-4t^2/U$, as expected from the large $U$
expansion.\cite{Mol93b} For the DW sites we obtain that $E_t^y$ closely 
reproduces one half of the kinetic energy in the PM phase at the actual 
hole doping found at these sites. In the two cases shown in Fig. 
\ref{fig:Ekin16}, the kinetic energy is weakly influenced by the 
involved band structure describing the stripe phases, and its profile 
mostly interpolates between the two expected limits. For the FVS its 
most significant effect is seen in $E_t^x$ close to the DW's, where most 
of the kinetic energy gain is realized. The same trends are also found 
for smaller $U$. 
 
As expected, profile of $E_t^x(l_x)$ and $E_t^y(l_x)$ depends on 
whether we include local electron correlations or not, as we have 
already seen that they markedly influence the charge and spin 
distribution. On the one hand, one sees that the AF domains between 
stripes are characterized by a more favorable kinetic energy gain in the 
SBA, owing to a combined effect of a weaker spin polarization and larger 
doubly occupancies than those found in the HFA. On the other hand, in 
the SB approach transverse charge fluctuations on the bonds connecting 
the DWs, with a strong reduction of $D$, and their nearest neighbors, 
are less efficient than the corresponding ones found in the HFA, where 
the fact that DWs are nonmagnetic implies much larger $D$ at these sites 
[\textit{cf}. Fig.~\ref{fig:Ekin16}(a)]. However, in the SBA, a small 
$D$ does not necessarily imply a strong reduction of the band narrowing 
factor. Indeed, the operator $z_{\sigma\sigma'}$ describes a sum of 
the two possible transition channels, either of which may accompany 
an electron hopping process:\cite{Fre92}  
(singly occupied site) $\rightarrow$ (empty site), or 
(doubly occupied site) $\rightarrow$ (singly occupied site 
                                      with a time-reversed spin). 
Therefore, as the hole density $n_h$ at DW sites is large, the first 
channel should even dominate, which follows from a simple relation,
\begin{equation}
e^2 = n_h + d^2.
\label{eq:e}
\end{equation}
that follows for a PM site from the local constraint (\ref{eq:const1}).
In addition, the HF stripe phase has a twice smaller electron density 
at the DWs than predicted by the SBA. Altogether, despite large double 
occupancies at the DWs, the HF kinetic energy gain along the FVS is 
smaller than the SB one, as shown in Fig.~\ref{fig:Ekin16}(b).  

\begin{table}[b!]
\caption 
{
Comparison of the site-normalized kinetic energy contribution for bonds 
along (10) ($E_t^{x}$) and (01) ($E_t^{y}$) directions, as obtained for 
the phase with either the FVS or HVS in the Hubbard model on an 
$128\times 128$ cluster at $x=1/16$ doping and $\beta t=1000$ 
within the HFA and SBA. 
}
\label{tab:Ekinx1_16}
\begin{ruledtabular}
\begin{tabular}{cccccccc}
\multicolumn{2}{c}{}               &   &
\multicolumn{2}{c}{FVS}            &
\multicolumn{1}{c}{}               &
\multicolumn{2}{c}{HVS}           \\
\multicolumn{1}{c}{$U/t$}                &
\multicolumn{1}{c}{method}          &   &
\multicolumn{1}{c}{$E_{t}^{x}/t$}         &
\multicolumn{1}{c}{$E_{t}^{y}/t$}         &
\multicolumn{1}{c}{}                      &
\multicolumn{1}{c}{$E_{t}^{x}/t$}         &
\multicolumn{1}{c}{$E_{t}^{y}/t$}         \\
\colrule
 6 & HFA & &  $-$0.5782  & $-$0.5436  & & $-$0.5337  & $-$0.5905  \\
   & SBA & &  $-$0.6085  & $-$0.5844  & & $-$0.5638  & $-$0.6171  \\
12 & SBA & &  $-$0.3872  & $-$0.3583  & & $-$0.3562  & $-$0.3794  \\
\end{tabular}
\end{ruledtabular}
\end{table}
 
Nevertheless, we have found a confirmation of the trend detected in the
earlier calculations\cite{Zaa96,Nor01} performed in the HFA --- also in 
the SBA the FVS are mainly stabilized not by the charge motion along the 
DWs but rather by the hole delocalization along the bonds perpendicular 
to the stripes, i.e., by the so-called solitonic mechanism. This effect 
is well illustrated by both components, $E_{t}^{x}$ and $E_{t}^{y}$, of 
the kinetic energy $E_t$, shown in Table \ref{tab:Ekinx1_16}. Here, one 
observes a significant anisotropy between the HF kinetic energy gain in 
the $x$ and $y$-directions, $\delta E=E_{t}^{x}-E_{t}^{y}$. This 
anisotropy is considerably reduced when the local electron correlations 
are included in the SBA, but remains the driving mechanism stabilizing 
the stripes.

These results should be now compared with those in Figs.
\ref{fig:Ekin16}(c) and \ref{fig:Ekin16}(d), obtained for the structure 
with the HVS. First of all, note that as in the case of the filled 
stripes, inclusion of the strong electron correlations improves the 
kinetic energy gain in the AF domains. However, the fact that the 
half-filled DWs are characterized by a smaller hole density 
than their filled counterparts results in an entirely different 
mechanism being responsible for their stability. Indeed, Fig. 
\ref{fig:Ekin16}(d) shows that the largest kinetic energy gain is 
released by the electrons propagating along the half-filled DWs,
while the transverse charge fluctuations are less important. This effect 
is particularly strong in the HFA, where one finds a large anisotropy 
between $E_t^x(l_x)$ and $E_t^y(l_x)$ at the DWs. We ascribe this to the 
fact that the increase of electron density at the nonmagnetic vertical 
DWs raises double occupancy (\textit{cf}. Figs. \ref{fig:nszdF16} and 
\ref{fig:nszdH16}) in this approach, and therefore facilitates the 
on-wall propagation of the electrons. Moreover, the nearest neighbor 
sites of the half-filled DW possess a much less quenched $S_{\pi}$, as 
compared to the neighbors of the filled one. Therefore, all the 
electrons crossing HVS encounter in this case a stronger on-site 
potential that develops in the AF domains, which acts to suppress 
locally $E_t^x(l_x)$. A similar amplification (reduction) of the 
on-wall (transverse) kinetic energy gain, respectively, we have also 
found in the SBA.  

We conclude the above analysis by showing in Table~\ref{tab:Ekinx1_16} 
two kinetic energy components $E_{t}^{x}$ and $E_{t}^{y}$ as found for 
the phase with the HVS. As in the case of the FVS, the presence of the 
HVS introduces a finite anisotropy between the kinetic energy gain in 
the $x$ and $y$ directions but while $E_{t}^{x}<E_{t}^{y}$ for the phase 
with the FVS, the structure with the HVS is characterized instead by 
$E_{t}^{x}>E_{t}^{y}$.

\begin{figure}[t!]
\begin{center}
\includegraphics*[width=8.2cm ]{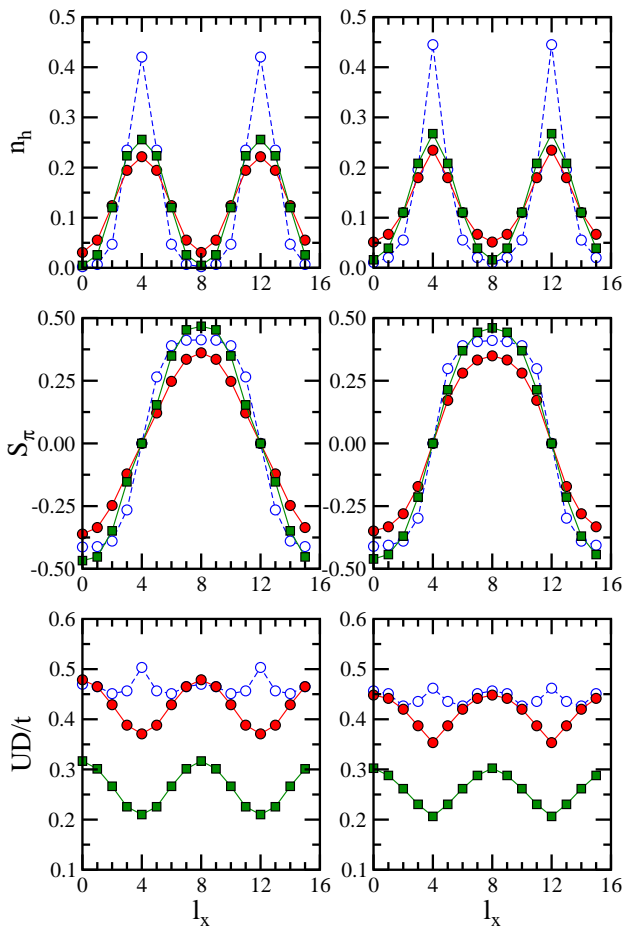}
\end{center}
\caption
{(Color online)
Local hole $n^{}_{\rm h}(l_x)$ (top) and magnetization $S_{\pi}(l_x)$ 
(second row) density, as well as double occupancy $D(l_x)$ (bottom), 
of the FVS (left) and FDS (right) phase found in the Hubbard model on 
an $128\times 128$ cluster for doping $x=1/8$ at $\beta t=1000$. 
Open (filled) circles show the HF (SB) results for $U=6t$, respectively, 
while filled squares denote the SB data for $U=12t$. 
}
\label{fig:nszdF}
\end{figure}

Let us now discuss variation of the kinetic energy gain shown in Figs. 
\ref{fig:Ekin16}(e) and \ref{fig:Ekin16}(f) across the unit cell of the 
phase with either the FDS or HDS. Here one finds that due to the 
symmetry $E_{t}^{}/2=E_{t}^{x}=E_{t}^{y}$. Therefore, based on Fig. 
\ref{fig:Ekin16}(e) showing substantial kinetic energy gain due to 
transverse hopping processes onto and off the stripe, one might expect 
that such a structure enables the largest kinetic energy gain among the 
stripe phases. In fact, it happens only in the strong coupling regime 
with $U=12t$, as reported in Table~\ref{tab:Ftp0x1_16}. In contrast, 
a weak charge modulation induced by the HDS results in a smooth kinetic 
energy profile without a particularly severe gain in the vicinity of 
the DWs. As a consequence, this phase is characterized by the smallest 
kinetic energy gain regardless of both the approach and the strength of 
$U$ (see Table~\ref{tab:Ftp0x1_16}).

\subsection{ Stripes at intermediate doping $x=1/8$ }
\label{sec:3c}
  
\begin{figure}[t!]
\begin{center}
\includegraphics*[width=8.2cm ]{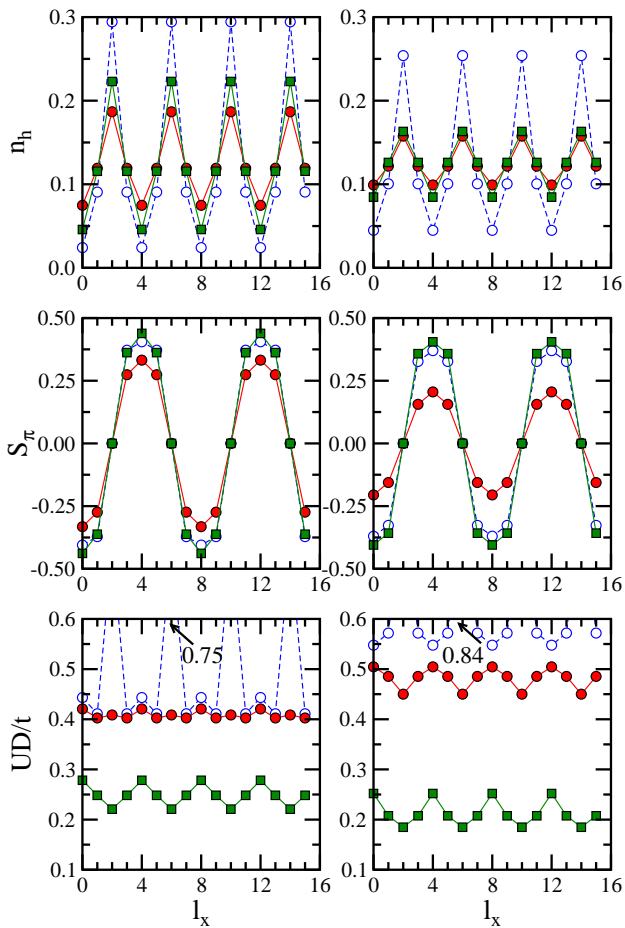}
\end{center}
\caption
{(Color online)
The hole density $n^{}_{\rm h}(l_x)$ (top), magnetization density 
$S_{\pi}(l_x)$ (second row), and double occupancy $D(l_x)$ (bottom) 
for the half-filled stripe phases: HVS (left) and HDS (right).
Open (filled) circles show the HF (SB) results for $U=6t$, respectively, 
while filled squares denote the SB data for $U=12t$.
Doping, cluster size and temperature as in Fig. \ref{fig:nszdF}. 
}
\label{fig:nszdH}
\end{figure}

\begin{figure}[t!]
\begin{center}
\includegraphics*[width=8.2cm ]{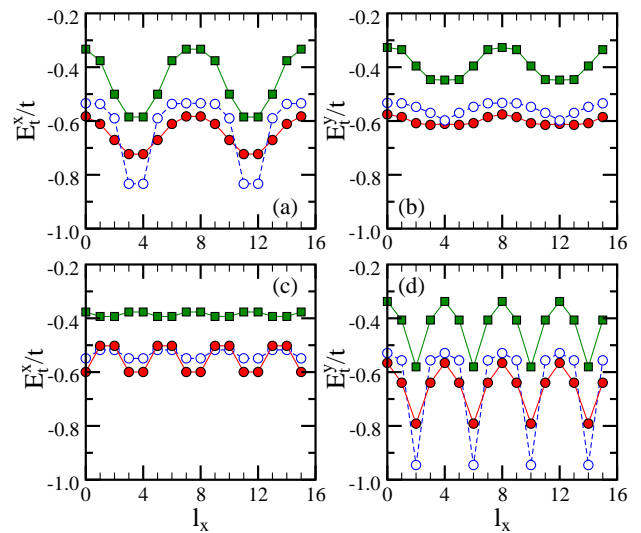}
\end{center}
\caption
{(Color online)
Kinetic energy $E_{t}^{x}(l_x)$ and $E_{t}^{y}(l_x)$  projected on the 
bonds along the $x$ and $y$ direction, as obtained for the phase with 
either the FVS (a,b), or HVS (c,d), shown in Figs.~\ref{fig:nszdF} and 
\ref{fig:nszdH}, respectively. 
Open (filled) circles show the HF (SB) results for $U=6t$, respectively, 
while filled squares denotes the SB data for $U=12t$.
}
\label{fig:Ekin}
\end{figure}

We turn now to a short discussion of the stripe profiles shown in 
Figs. \ref{fig:nszdF} and \ref{fig:nszdH}, obtained at a twice higher 
doping level $x=1/8$. The stripe structures at this doping confirm  
the main conclusions of the previous Section concerning the stripe 
stability. One finds again that the transverse (parallel) hopping 
stabilizes the structures with filled (half-filled) vertical DWs, 
respectively (\textit{cf}. Fig. \ref{fig:Ekin}). In addition, 
we point out the most prominent changes with respect to the low doping 
regime. 

First of all, the increased doping acts to shrink the distance between 
the filled DWs according to the formula $d=1/x$, so that the charge 
(magnetic) unit cell is reduced and involves now eight (sixteen) atoms, 
respectively, as depicted in Fig. \ref{fig:nszdF}. One finds that the
charge distribution is smoother in the SBA than in the HFA, and in the
moderate coupling regime $U=6t$ the highest hole density on the 
nonmagnetic ($S_{\pi}=0$) DW atoms is only $n_h\simeq 0.22\div0.24$. 
This demonstrates that the filled stripe phases
are truly itinerant systems with modulated charge density.  

\begin{table}[b!]
\caption 
{
The same as in Table~\ref{tab:Ftp0x1_16} but for $x=1/8$. 
}
\label{tab:Ftp0x1_8}
\begin{ruledtabular}
\begin{tabular}{cccccc}
\multicolumn{1}{c}{method}        &
\multicolumn{1}{c}{$U/t$}         &
\multicolumn{1}{c}{phase}         &
\multicolumn{1}{c}{$E_U/t$}       &
\multicolumn{1}{c}{$E_t/t$}       &
\multicolumn{1}{c}{$F/t$}         \\
\colrule
  HFA  & 6& HDS & 0.6321 & $-$1.2330 & $-$0.6009  \\
       &  &  AF & 0.5148 & $-$1.1808 & $-$0.6660  \\
       &  & HVS & 0.5031 & $-$1.1800 & $-$0.6769  \\
       &  & FVS & 0.4649 & $-$1.1777 & $-$0.7128  \\
       &  & FDS & 0.4432 & $-$1.1639 & $-$0.7207  \\ 
  SBA  & 6& HDS & 0.4815 & $-$1.2713 & $-$0.7898  \\
       &  &  AF & 0.4563 & $-$1.2563 & $-$0.8000  \\
       &  & HVS & 0.4083 & $-$1.2098 & $-$0.8014  \\
       &  & FVS & 0.4269 & $-$1.2470 & $-$0.8201  \\
       &  & FDS & 0.4123 & $-$1.2324 & $-$0.8201  \\ 
  SBA  &12& HDS & 0.2130 & $-$0.7467 & $-$0.5337  \\
       &  &  AF & 0.2726 & $-$0.8112 & $-$0.5386  \\
       &  & HVS & 0.2486 & $-$0.8175 & $-$0.5689  \\
       &  & FVS & 0.2645 & $-$0.8396 & $-$0.5751  \\
       &  & FDS & 0.2588 & $-$0.8409 & $-$0.5821  \\ 
\end{tabular}
\end{ruledtabular}
\end{table}

However, the experimentally established distance between stripes in 
Nd-LSCO at $x=1/8$ doping is equal to $d=4$,\cite{Tra95Nd} implying that 
{\it de facto\/} less holes are doped per one DW site, and consequently 
DWs are only partly filled by holes. Indeed, other local minima are 
obtained in the SB calculations for the charge (magnetic) unit cells of 
the half-filled stripe phases, containing four (eight) atoms, 
respectively, as shown in Fig. \ref{fig:nszdH}. 
Also these phases are itinerant, with the SB hole 
densities at $U=6t$ less than 0.2 at the DW sites for both half-filled 
stripe phases. Moreover, correlations implemented in the SB approach 
guarantee that electrons avoid each other and the double occupancy is 
the lowest $D\simeq 0.07$ at the half-filled vertical DW sites, while in 
the HFA one finds instead a large value $D_0\simeq 0.13$. However, it is 
remarkable that the SB double occupancy $D$ is practically site 
independent in the phase with the HVS, in spite of rather different 
electron densities at the DW sites and within the AF domains. 

Next, we address a generic crossover found in the SBA from the FDS to 
their vertical counterparts taking place upon doping. We have seen from
the respective charge distributions in Fig.~\ref{fig:nszdF} that a 
certain amount of holes proliferates out of the DWs, reducing double 
occupancies in the AF domains. Consequently, the on-site energy gains 
become less important, which explains the transition towards the phase 
with the FVS, being a more favorable structure for the charge dynamics. 
In fact, for $U=6t$, this crossover occurs precisely at $x=1/8$, so that 
both phases become degenerate, as reported in Table~\ref{tab:Ftp0x1_8}, 
while in the strong coupling regime $U=12t$ one recovers in the ground 
state the phase with the FDS.  
Moreover, as the doped holes are now only weakly localized at the 
DW sites, a nonuniform charge distribution in the AF domains amplifies 
the anisotropy between the kinetic energy gain $E_{t}^{x}$ and 
$E_{t}^{y}$ (\textit{cf}. Table~\ref{tab:Ekinx1_8}).

\begin{table}[b!]
\caption 
{
The same as in Table~\ref{tab:Ekinx1_16} but for $x=1/8$. 
}
\label{tab:Ekinx1_8}
\begin{ruledtabular}
\begin{tabular}{cccccccc}
\multicolumn{2}{c}{}               & &
\multicolumn{2}{c}{FVS}            &
\multicolumn{1}{c}{}               &
\multicolumn{2}{c}{HVS}           \\
\multicolumn{1}{c}{$U/t$}                &
\multicolumn{1}{c}{method}              & &
\multicolumn{1}{c}{$E_{t}^{x}/t$}         &
\multicolumn{1}{c}{$E_{t}^{y}/t$}         &
\multicolumn{1}{c}{}                      &
\multicolumn{1}{c}{$E_{t}^{x}/t$}         &
\multicolumn{1}{c}{$E_{t}^{y}/t$}         \\
\colrule
 6 & HFA & &  $-$0.6234  & $-$0.5543  & & $-$0.5331  & $-$0.6469 \\
   & SBA & &  $-$0.6468  & $-$0.6002  & & $-$0.5504  & $-$0.6593   \\
12 & SBA & &  $-$0.4486  & $-$0.3910  & & $-$0.3848  & $-$0.4327   \\
\end{tabular}
\end{ruledtabular}
\end{table}
\subsection{ Electronic structure of stripe phases }
\label{sec:3d}

Let us analyze in more detail the electronic structure obtained for the 
stripe phases in the low doping regime. Since the stripe phases consist 
of a superposition of magnetic domains with a 2D character, and domain 
walls of 1D character, it is worth determining to which part of the 
spectrum each of them contribute, and to which extend they mix. The DOS 
of a pure 2D AF phase exhibits a gap ($\Delta\sim U$) separating two 
van Hove singularities lying at the top (bottom) of the lower (upper) 
Hubbard subbands, here representing the dynamical Hubbard bands. In 1D 
case the van Hove singularities are located at the lower and upper band 
edges.

\begin{figure}[t!]
\begin{center}
\includegraphics*[width=8.2cm ]{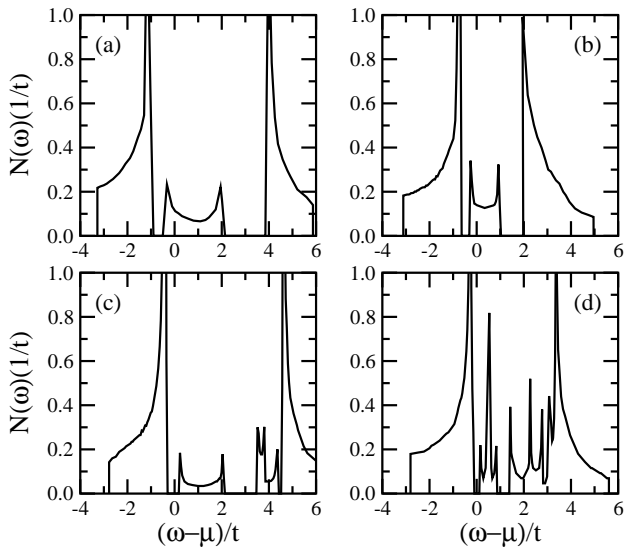}
\end{center}
\caption
{
Density of states $N(\omega)$ of the phase with: 
(a,b) HVS, shown in the left panels of Fig.~\ref{fig:nszdH16}; 
(c,d) FVS, depicted in the left panels of Fig.~\ref{fig:nszdF16}, 
as obtained within the HFA (left) and the SBA (right) in the moderate
coupling regime $U=6t$ for doping $x=1/16$. 
}
\label{fig:DOSHFSB16}
\end{figure}

Let us consider the actual DOS for the phase with the HVS in the 
moderate coupling regime $U=6t$ --- it consists indeed of two Hubbard 
subbands at low and high energies, lower Hubbard band (LHB) and upper 
Hubbard band (UHB), separated by a gap of $\sim 5t$ in the HFA and 
$\sim 2.6t$ in the SBA, see Figs.~\ref{fig:DOSHFSB16}(a) and 
\ref{fig:DOSHFSB16}(b). So strong reduction of the gap with a similar 
value of the magnetization in the AF domains is the first correlation 
effect identified in the electronic structure.   

Furthermore, the formation of stripes with nonmagnetic sites results 
in additional electronic states within the Mott-Hubbard gap. These new 
structures contain parts of the spectral density below the Fermi energy, 
both in the HFA and in the SBA, indicating that half-filled stripe 
phases are metallic. Note also that the nonmagnetic stripes hybridize 
only weakly with the neighboring AF domains, since they are mainly 
stabilized by the electrons propagating along them, so this part of 
$N(\omega)$ is reminiscent of the tight binding DOS for a linear chain 
with nearest neighbor hopping, and has the characteristic peaks at the 
edges. We recall that in 1D case the van Hove singularities are located 
at the lower and upper band edges, as found here for the in-gap bands. 
Similar to the Hubbard subbands, also the partly filled electronic 
subbands in the gap are substantially narrowed by the correlation 
effects in the SBA. 

In the case of filled vertical stripe phases shown in Figs. 
\ref{fig:DOSHFSB16}(c) and \ref{fig:DOSHFSB16}(d), the states in the gap 
are empty. This was actually pointed out rather early as the microscopic 
reason responsible for the stability of such phases in the HFA.
\cite{Zaa96} The most prominent feature of the DOS for FVS phase is 
again a striking similarity of the mid-gap part of $N(\omega)$ to the 
DOS of a linear chain. The width is somewhat reduced indicating a more 
localized character of these states. In contrast with the phase with the 
HVS, a stronger hybridization with the AF domains influences the UHB so 
that a certain amount of its states is shifted towards a lower energy 
and forms a separate segment of $N(\omega)$ 
[\textit{cf}. Fig. \ref{fig:DOSHFSB16}(c)].

Finally, in Fig.~\ref{fig:DOSHFSB16}(d) we show the DOS of the phase 
with the FVS which follows from the SBA. In this case the charge and 
spin density profiles are more spread out and a weaker anisotropy 
between the tranverse and parallel electron motion, as compared to the 
HF ones, results in a less clear character of the mid-gap part of 
$N(\omega)$, with a large peak in the middle being now a characteristic 
feature of the 2D tight binding DOS. Notice also an insulating nature of 
both filled stripes. Indeed, as the mid-gap states are now entirely 
empty, the Fermi energy lies inside the gap between the highest occupied 
state of the LHB and the bottom of the mid-gap band.

\begin{figure}[t!]
\begin{center}
\includegraphics*[width=8.2cm ]{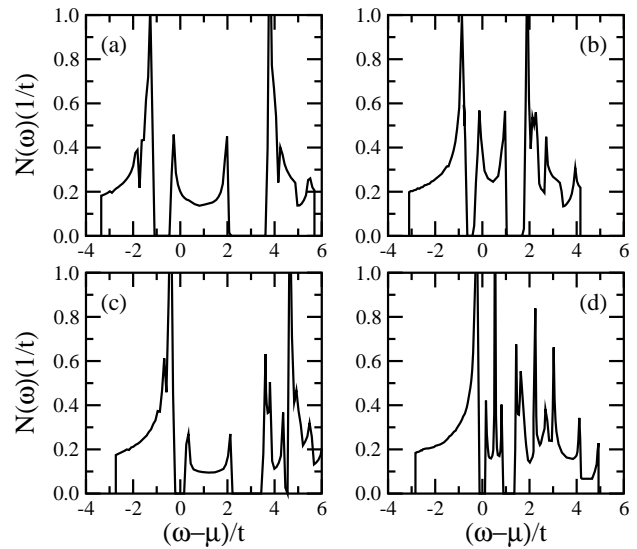}
\end{center}
\caption
{
Density of states $N(\omega)$ of the phase with: 
(a,b) HVS, shown in the left panels of Fig.~\ref{fig:nszdH}; 
(c,d) FVS, depicted in the left panels of Fig.~\ref{fig:nszdF}, 
as obtained within the HFA (left) and SBA (right) in the moderate
coupling regime $U=6t$ for doping $x=1/8$. 
}
\label{fig:DOSHFSB}
\end{figure}

For the higher doping $x=1/8$ one finds similar features in the DOS of 
both HVS and FVS phases. The spectral intensity is here redistributed, 
showing a systematic transfer of the spectral weight from the UHB into 
these states and the LHB upon doping, in agreement with a strong 
coupling perturbation theory for the Hubbard model.\cite{Esk94} 
In addition, the spectral weight is also transfered into the mid-gap 
part of $N(\omega)$, which grows with an increasing number of stripes 
in the cluster.

\section{\label{sec:4} Stripe phases in the extended hopping model }

\subsection{ Crossover to vertical stripe phases }
\label{sec:4a}

Thus far we have worked out that in the strongly correlated regime
relevant for the cuprates a structure with the FDS (HVS) is the lowest 
in energy among the stripe structures with the filled (half-filled) DWs, 
respectively. Therefore, an interesting question occurs --- which 
microscopic parameters decide whether the phase with the FDS or the one 
with the HVS is more stable. To clarify this point we have investigated 
in the SBA the competition between the stripe phases in the $t$-$t'$-$U$ 
model with $U=12t$. The chosen value of the Coulomb interaction gives 
the ratio $J/t=1/3$ which corresponds to the physical value in the 
cuprates.\cite{Jef92}

\begin{figure}[t!]
\begin{center}
\includegraphics*[width=8.2cm ]{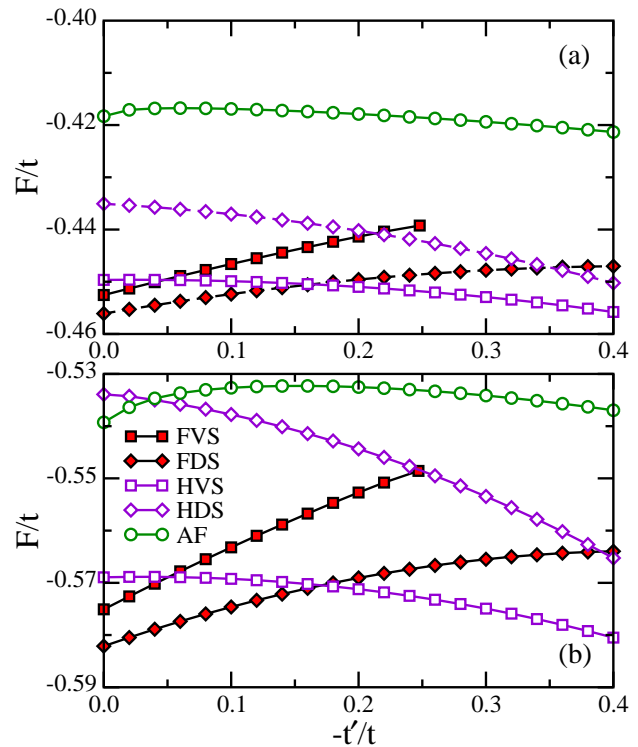}
\end{center}
\caption
{(Color online) 
Site-normalized SB free energy $F$ of the phase with the FVS, FDS, 
HVS, and HDS, as well as of the AF phase, as a function of increasing 
next-nearest neighbor hopping $|t'|$ for $U=12t$ in the: 
(a) underdoped regime $x=1/16$; (b) moderately doped regime $x=1/8$.
}
\label{fig:F}
\end{figure}

\begin{figure}[t!]
\begin{center}
\includegraphics*[width=8.2cm ]{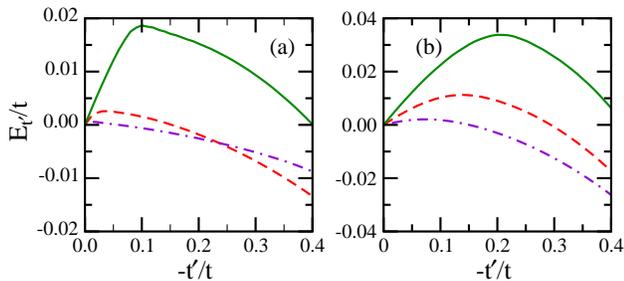}
\end{center}
\caption
{(Color online) 
Average kinetic energy $E_{t'}$ per diagonal bond as a function of 
increasing next-nearest neighbor hopping $|t'|$ for:
(a) AF phase, and 
(b) PM phase, 
as obtained for $U=12t$. Dot-dashed, dashed, and solid line shows the 
results for $x=1/16$, $x=1/8$, and $x=1/5$ doping, respectively.
}
\label{fig:AF}
\end{figure}

Fig.~\ref{fig:F}(a) shows the effect of increasing next-nearest neighbor 
hopping $|t'|$ on the free energy $F$ of all four stripe phases, as well 
as of the uniform AF phase at the hole doping $x=1/16$ and $U=12t$. 
Here, the most striking result is that while increasing $|t'|$ only 
weakly influences the energy of the AF phase, it modifies substantially 
$F$ of the stripe phases. Moreover, it clearly stabilizes half-filled 
stripes reducing simultaneously stability of the filled DWs. To 
appreciate better the enhanced stability of the half-filled stripes with 
respect to the latter ones, we analyze the average kinetic energy 
$E_{t'}$ per diagonal bond as a function of increasing $|t'|$, obtained 
in the AF and PM phases for a few doping levels: $x=1/16$ 
(dot-dashed line), $x=1/8$ (dashed line), and $x=1/5$ (solid line) (see 
Fig. \ref{fig:AF}). One finds that a negative next-nearest neighbor 
hopping ($t'/t<0$) yields first a positive kinetic energy contribution 
with the loss being proportional to doping. On the one hand, hopping 
processes associated with $t'$ are partly suppressed in the AF phase by 
both the reduction of double occupancies and the increase of spin 
polarization. On the other hand, in the PM phase, these processes might 
be optimized only by the first effect, which results in a much more 
severe loss of the energy. However, these trends also involve a 
reduction of the nearest neighbor hopping processes and therefore they 
cannot be maintained above a certain critical value of $t'$. 
As a result, $E_{t'}$ changes its sign and becomes negative 
(\textit{cf}. Fig.~\ref{fig:AF}).     

Turning back to the analysis of the stripe stability, the fact that the 
energy loss due to the $t'$ term follows predominately from the hopping 
processes at the nonmagnetic DWs explains immediately why a finite $t'$ 
promotes partially filled stripes. Indeed, as the average hole density 
at the half-filled DWs is smaller than at the filled ones 
(\textit{cf}. Figs.~\ref{fig:nszdF16} and \ref{fig:nszdH16}), 
one should lose less kinetic energy in the former case, especially in 
the phase with the HDS, where the reduction of the hole density at the 
DWs is the strongest. Consequently, as depicted in Fig.~\ref{fig:F}(a), 
this phase is characterized by the largest free energy gain upon 
increasing $|t'|$. Further, increasing $|t'|$ strongly interferes with 
the solitonic mechanism\cite{Zaa96} stabilizing the FVS by the 
transverse hopping and hence such a structure becomes unstable at 
$t'/t\simeq -0.248$. In contrast, narrower FDS are stabilized to a 
lesser degree by this mechanism remaining a stable solution in the 
entire region $0<|t'/t|<0.4$. Finally, the energy difference between the 
phase with the HVS and the one with the FDS gradually diminishes so that 
the former becomes the ground state of the system at $t'/t\simeq -0.16$.

In order to get a more quantitative insight into the mechanism of the 
crossover between the above phases, we show in Table~\ref{tab:Ftp} 
decomposition of their free energy into the on-site Coulomb energy $E_U$ 
as well as into both kinetic energy contributions $E_{t}$ and $E_{t'}$.
On the one hand, the robust stability of the FDS in the absence of $t'$, 
follows mainly from a large kinetic energy gain $E_t$ 
(\textit{cf}. Table~\ref{tab:Ftp0x1_16}). 
In contrast, at the expense of $E_t$, the structure with the HVS better 
optimizes the double occupancy energy $E_U$, being however less stable 
than the one with the FDS. On the other hand, as we have already found, 
a negative next-nearest neighbor hopping tends to yield a positive 
kinetic energy contribution, but the actual sign depends both on the 
value of $t'$ and on the filling of the DWs. Indeed, the presence of the 
half-filled stripes clearly helps to alterate the disadvantageous sign 
of $E_{t'}$ so that it is negative already for $t'=-0.3t$ (\textit{cf}. 
Table~\ref{tab:Ftp}). Conversely, in the phase with the FDS, in spite of 
a stronger reduction of double occupancies with a concomitant detriment 
of $E_t$, delocalization of electrons due to a finite $t'=-0.3t$ still 
costs a small amount of energy. All these features act to stabilize the 
HVS in the ground state.
 
\begin{table}[b!]
\caption 
{
Free energy $F$ per site, on-site Coulomb energy $E_U$ as well as 
kinetic energy $E_t$ and $E_t'$ contributions, as obtained for the phase 
with either the HVS or FDS in the extended Hubbard model from the SBA 
on an $128\times 128$ cluster.
Parameters: $U=12t$, $t'=-0.3t$ and $\beta t=1000$.
}
\label{tab:Ftp}
\begin{ruledtabular}
\begin{tabular}{ccccrc}
\multicolumn{1}{c}{$x$}           &
\multicolumn{1}{c}{phase}         &
\multicolumn{1}{c}{$E_U/t$}       &
\multicolumn{1}{c}{$E_t/t$}       &
\multicolumn{1}{c}{$E_{t'}/t$}    &
\multicolumn{1}{c}{$F/t$}         \\
\colrule
$1/16$    & FDS & 0.2775 & $-$0.7292 &    0.0039 & $-$0.4478 \\ 
          & HVS & 0.2767 & $-$0.7225 & $-$0.0071 & $-$0.4529 \\    
$1/8$     & FDS & 0.2332 & $-$0.8064 &    0.0077 & $-$0.5655 \\ 
          & HVS & 0.2300 & $-$0.7911 & $-$0.0138 & $-$0.5749 \\   
\end{tabular}
\end{ruledtabular}
\end{table}

Based on the above discussion one could expect that at a higher doping, 
the HVS would take over the FDS for a much smaller value of $|t'|$. 
Indeed, as shown in Table~\ref{tab:Ftp}, $x=1/8$ provides a more 
significant energy lowering of the phase with the HVS due to a twice 
larger $E_{t'}$ gain. Simultaneously, the energy loss in the structure 
with the FDS due to the $t'$ processes is twice as large as for 
$x=1/16$. Consequently, increasing $|t'|$ destabilizes now easier the 
FVS and one finally loses this solution already at $t'/t\simeq -0.247$, 
as illustrated in Fig.~\ref{fig:F}(b). 
Nevertheless, one reads off from Fig.~\ref{fig:F}(b) that a change 
of the ground state upon increasing $|t'|$ takes place in this case 
at even slightly larger value $|t'|\simeq 0.17t$. The explanation is 
contained in Table~\ref{tab:Ftp}: a higher doping diminishes the 
double occupancies and additionally it is reflected in a higher 
mobility of the electrons. As a consequence, the relative energy 
\begin{equation} 
\delta F=F_{\rm FDS}-F_{\rm HVS},
\label{eq:F}
\end{equation}
varies, in the absence of $t'$, from $\delta F=-0.0065t$ for $x=1/16$ 
to $\delta F=-0.0132t$ at $x=1/8$, which demonstrates an increasing 
stability of the FDS. Therefore, the condition of having the crossover 
at $x=1/8$ requires overcompensation of the larger $|\delta F|$ and 
thus the transition is now slightly moved towards the larger $|t'|$. 

\begin{figure}[t!]
\begin{center}
\includegraphics*[width=8.2cm ]{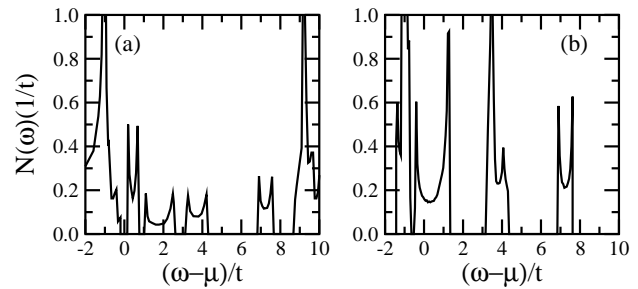}
\end{center}
\caption
{
Density of states $N(\omega)$ of the ground state with: 
(a) FDS for $x=1/16$ and $t'=0$; 
(b) HVS for $x=1/8$ and $t'=-0.3t$, 
as obtained for large Coulomb interaction $U=12t$.   
}
\label{fig:DOSgs}
\end{figure}

Altogether, Fig.~\ref{fig:F} enables scenario in which doping-induced 
increase in the amplitude of $t'$ results in a drastic change in the 
spin modulation from the diagonal to vertical/horizontal one, as 
observed experimentally in LSCO at $x\simeq 0.06$. The conjecture that 
$|t'|$ is suppressed in underdoped regime and grows with increasing $x$, 
seems to be also supported by the recent ED studies showing that $t'$ 
helps to stabilize the superconductivity in the optimally and overdoped 
regime but it is harmful in the underdoped regime.\cite{Shi04} 
Moreover, inelastic neutron scattering experiments on YBCO, a bilayered 
compound with a larger than LSCO next-nearest neighbor hopping term 
$|t'|\simeq 0.3t$,\cite{Pav01} have established the presence of the 
IC {\it vertical/horizontal\/} spin fluctuations throughout its entire 
superconducting regime,\cite{Dai01} implying that such a modulation is: 
 (i) realized easier for larger $|t'|$, and  
(ii) more advantageous for the superconductivity than the diagonal one.
Note, however, that it remain puzzling why the average filling of 
diagonal DWs changes from one to $(1/\sqrt{2})$ hole per one DW site
in the very low doping regime $x\sim 0.03$. Unfortunately, we could not
address this issue here as it requires calculations for still larger 
unit cells which could not be performed at present.   

\subsection{ Changes in the electronic structure due to finite $t'$ }
\label{sec:4b}

Remarkably, in the present approach, the two possible ground states have 
entirely different physical properties being either insulating or 
metallic. Indeed, the DOS of the ground state with the FDS obtained in 
the Hubbard model with $t'=0$ at $x=1/16$ consists of clearly seen two 
distinct maxima corresponding to the Hubbard bands, while the Fermi 
energy falls within a gap which results from the magnetic order 
[\textit{cf}. Fig.~\ref{fig:DOSgs}(a)]. Interestingly, due to a strong 
Coulomb repulsion $U=12t$ and consequently small band narrowing factors, 
instead of the ones as for $U=6t$, one finds here four separated midgap 
bands. In contrast, as shown in Fig.~\ref{fig:DOSgs}(b), the Fermi 
energy crosses the midgap states of the structure with the HVS, which is 
the ground state of the Hamiltonian with $x=1/8$ and $t'=-0.3t$, 
enabling charge transport in agreement with the data for LSCO. 

\begin{figure}[t!]
\begin{center}
\includegraphics*[width=8.2cm ]{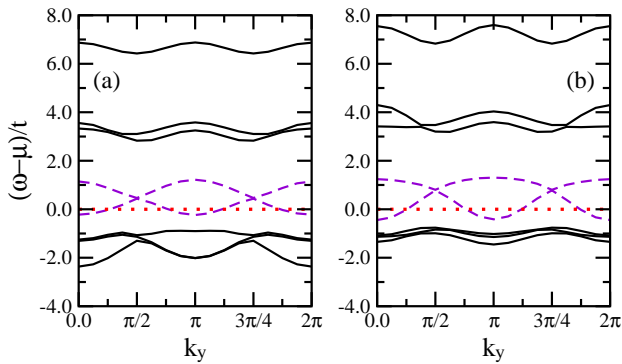}
\end{center}
\caption
{(Color online) 
Electronic structure of the phase with the HVS as a function of parallel 
momentum $k_y$, calculated from the unit cell shown in Fig. 
\ref{fig:unit}(a) on an 8$\times$8 cluster within the SB method for 
$U/t=12$, and: 
(a) $t'=0$, 
(b) $t'/t=-0.3$. 
Solid (dashed) lines correspond to the bulk (mid-gap) bands, 
respectively, whereas the dotted lines indicate the Fermi level. 
}
\label{fig:single}
\end{figure}

We also note that, in agreement with Ref.~\onlinecite{Sei04}, 
a negative $t'$ leads to a distinct broadening of the partially filled 
mid-gap band induced by the HVS and consequently it shifts these states 
to a lower energy, as depicted in Fig.~\ref{fig:single}. Conversely, 
the energy gain due to such a modification of the electronic structure 
is not possible in the case of the FDS as their mid-gap states are 
entirely \emph{unoccupied}. Thus, the puzzling role of $t'$ in promoting 
the half-filled DWs is clarified.

\section{ Discussion and Summary }
\label{sec:5}

We have developed a simple but powerful SB approach which allows one to 
investigate various stripe phases with a large unit cell and carry out 
the calculations on a large ($128\times 128$) cluster. Having presented 
the theoretical framework in Sec.~\ref{sec:2}, we have shown in Sec. 
\ref{sec:2d} that our method provides a unique opportunity to obtain 
unbiased results at low temperature $\beta t=1000$, as well as to 
eliminate the role of finite size effects. The electron correlation
effects beyond the HF were implemented within the SB method. Therefore, 
the stripe phases found in the present approach are stabilized not due 
to particular boundary conditions, but they represent a generic 
instability of the strongly correlated electron system doped by holes.

Then, in Sec.~\ref{sec:3} we have analyzed the stability of the 
idealized filled as well as the half-filled stripes in the Hubbard model 
at two representative doping levels $x=1/16$ and $x=1/8$. However, the 
true ground state could correspond to neither filled nor half-filled 
stripes, as the optimal filling might vary with doping. By comparing 
the SB charge and spin density profiles with the ones obtained in the 
HFA in moderate coupling regime with $U=6t$, we have emphasized the role 
of a proper treatment of electron correlations. In particular, we have 
found that the largest correction of the HF value is obtained at the DW 
unpolarized sites, where the double occupancy $D$ shows distinct minima, 
which allows one to optimize the on-site Coulomb energy $E_U$ even 
{\it without\/} a great reduction of the actual electron density. 
Simultaneously, an enhanced value of $D$ in SB approach suppresses 
partially the spin polarization of the atoms within the AF domains, 
which enables more intersite excitations and leads to a more favorable 
kinetic energy gain. Taken together, these two effects are responsible 
for smoother SB charge and spin density profiles with respect to the 
ones found in the HFA. Moreover, we have shown that at strong coupling 
with $U=12t$ the SB double occupancy profile is interpolating between 
$4t^2/U^2$ deep in the magnetic domains and the value expected in the 
PM phase at the hole density obtained at the DW sites.

However, the most prominent result of our extensive studies of the 
kinetic energy was a demonstration of a close relationship between the 
filling and the direction of the largest kinetic energy gain. Indeed, 
we have shown that while deep within the magnetic domains the kinetic 
energy in both directions is given by $-4t^2/U$, as expected from large 
$U$ expansions,\cite{Mol93b} in the neighborhood of DWs the physical 
situation is much more complex and depends on the microscopic 
properties of the considered stripe phase. Especially, it turns out that 
the HVS are predominately stabilized by the electrons propagating along 
them, whereas their filled counterparts might be considered as a set of 
weakly coupled solitons as their stability follows to a large extent 
from the transverse hopping across them.  

Finally, proceeding from the experimental motivation, based on the 
doping-induced change of the spatial orientation of the DWs in the 
cuprates, we have investigated the effect of the effective next-nearest 
neighbor hopping $t'$ on the stripe stability. Remarkably, we have found 
that a subtle increase of the amplitude of $t'$ with increasing doping 
could explain the observed transition from the filled insulating 
diagonal stripes towards the half-filled metallic vertical/horizontal 
ones in the cuprates. As the effective hopping $t'$ is different for 
hole-doped and electron-doped cuprates,\cite{Fei96} and the doped holes 
delocalize the magnetic moments at Cu sites, it might be expected that 
$|t'|$ increases with doping. This poses an interesting physical problem
for the parameters of the effective one-band model for the cuprates 
which should be resolved by future studies. Such a doping dependent 
$t'$ may also result from the change of lattice parameters upon doping 
observed in most superconducting cuprates.\cite{Ngu81}

\begin{acknowledgments}
We thank P. Dai, J. Spa\l{}ek, and B. Mercey for valuable discussions. 
M. R. acknowledges support from the European Community under 
Marie Curie Program Grant number HPMT2000-141. 
This work was supported by the Polish Ministry of Science and 
Education under Project No. 1~P03B~068~26, and by the Minist\`ere
Fran\c{c}ais des Affaires Etrang\`eres under POLONIUM 09294VH.
\end{acknowledgments}


\end{document}